\pdfoutput=1
\documentclass[aps,prx,reprint,floatfix,superscriptaddress,amsmath,amssymb,nofootinbib]{revtex4-2}
\usepackage[T1]{fontenc}
\usepackage[utf8]{inputenc}
\usepackage{lmodern,microtype,graphicx,booktabs,array,bm,xcolor,placeins,amsthm,enumitem,needspace}
\PassOptionsToPackage{hyphens}{url}
\usepackage[colorlinks=true,linkcolor=blue!55!black,citecolor=blue!55!black,urlcolor=blue!55!black]{hyperref}
\newtheorem{proposition}{Proposition}
\theoremstyle{remark}
\newtheorem*{remark}{Remark}
\newcommand{\leadin}[1]{\par\medskip\noindent\textbf{#1}\enspace\ignorespaces}
\newcommand{\proofstep}[1]{\par\smallskip\noindent\emph{#1}\enspace\ignorespaces}
\newcommand{\gate}[1]{\textnormal{\rmfamily #1}}

\newcommand{\ket}[1]{\lvert#1\rangle}
\newcommand{\bra}[1]{\langle#1\rvert}
\newcommand{\Qp}{Q_{\mathrm{peak}}}
\newcommand{\Ts}{T_{\mathrm{shot}}}
\newcommand{\Tf}{\widehat{T}_{\mathrm{fact}}}
\newcommand{\mb}{b_{\mathrm{mask}}}

\newcommand{\NumRefValid}{28}
\newcommand{\NumRefFrontier}{25}
\newcommand{\RefMinProdVal}{7.957\times 10^{12}}
\newcommand{\CmbNumConfigs}{49}
\newcommand{\CmbNumPareto}{30}
\newcommand{\CmbNumParetoEqBudget}{28}
\newcommand{\CmbMinQ}{1{,}306}
\newcommand{\CmbMaxQ}{2{,}919}
\newcommand{\CmbMinQTQ}{2{,}433}

\begin{document}
\title{Reducing RSA-2048 Factoring Resources Below One Billion Toffolis via LLM-Driven Verifiable Quantum Compilation}
\affiliation{Intelligent Quantum Inception Co., Ltd., Haidian, Beijing 100085, China}
\affiliation{Department of Physics and State Key Laboratory of Low Dimensional Quantum Physics, Tsinghua University, Beijing 100084, China}
\affiliation{Chen-Ning Yang Institute for Advanced Study, Tsinghua University, Beijing 100084, China}
\affiliation{Beijing Key Laboratory of Cold Atom Quantum Computation, Tsinghua University, Beijing 100084, China}
\author{Xiaodi Li}
\thanks{These authors contributed equally to this work.}
\affiliation{Intelligent Quantum Inception Co., Ltd., Haidian, Beijing 100085, China}
\affiliation{Beijing Key Laboratory of Cold Atom Quantum Computation, Tsinghua University, Beijing 100084, China}
\author{Yuqing Wang}
\thanks{These authors contributed equally to this work.}
\affiliation{Intelligent Quantum Inception Co., Ltd., Haidian, Beijing 100085, China}
\affiliation{Department of Physics and State Key Laboratory of Low Dimensional Quantum Physics, Tsinghua University, Beijing 100084, China}
\affiliation{Beijing Key Laboratory of Cold Atom Quantum Computation, Tsinghua University, Beijing 100084, China}
\author{Xue Chen}
\affiliation{Chen-Ning Yang Institute for Advanced Study, Tsinghua University, Beijing 100084, China}
\affiliation{Beijing Key Laboratory of Cold Atom Quantum Computation, Tsinghua University, Beijing 100084, China}
\author{Yang-Yang Tang}
\affiliation{Chen-Ning Yang Institute for Advanced Study, Tsinghua University, Beijing 100084, China}
\affiliation{Beijing Key Laboratory of Cold Atom Quantum Computation, Tsinghua University, Beijing 100084, China}
\author{Linghui Chen}
\email{chenlh@mail.ustc.edu.cn}
\affiliation{Intelligent Quantum Inception Co., Ltd., Haidian, Beijing 100085, China}
\affiliation{Beijing Key Laboratory of Cold Atom Quantum Computation, Tsinghua University, Beijing 100084, China}
\author{Tao Zhang}
\email{tao_zhang@mail.tsinghua.edu.cn}
\affiliation{Intelligent Quantum Inception Co., Ltd., Haidian, Beijing 100085, China}
\affiliation{Chen-Ning Yang Institute for Advanced Study, Tsinghua University, Beijing 100084, China}
\affiliation{Beijing Key Laboratory of Cold Atom Quantum Computation, Tsinghua University, Beijing 100084, China}
\date{November 22, 2026}

\begin{abstract}
Modular exponentiation dominates the logical cost of factoring 2048-bit RSA integers (RSA-2048) with Shor's algorithm. Recent approaches based on approximate residue number systems (RNS) sharply reduce the required logical qubits, yet still demand billions of Toffoli gates. Optimizing these circuits is difficult because shared ancilla qubits and measurement-based uncomputation tightly couple the arithmetic stages, while rare overflows in compressed registers occur too infrequently for random simulation to detect. Here we use large language model (LLM)-driven search to redesign the RNS pipeline across stages and build a deterministic compiler that produces a Pareto frontier of RSA-2048 circuits. We bound rare overflow probabilities analytically and verify the resulting elementary gate sequences across all constituent RNS primes on representative circuits. Relative to Gidney's 2025 baseline under matched adder accounting, our frontier cuts the minimum qubit--Toffoli product by 70.1\% and lowers expected factoring Toffolis by up to 78.6\% at matched accumulator precision, bringing RSA-2048 factoring below one billion Toffolis for the first time.
\end{abstract}

\maketitle

\let\realaddcontentsline\addcontentsline
\renewcommand{\addcontentsline}[3]{}

\section{Introduction}
\label{sec:intro}

Factoring 2048-bit RSA integers (RSA-2048) with Shor's algorithm~\cite{shor1997} remains an important application for fault-tolerant quantum computing. Over the past two decades, advances in reversible adders~\cite{cuccaro2004,gidney2018halving}, table lookups~\cite{babbush2018,low2024}, and Eker{\aa}--H{\aa}stad short discrete logarithms~\cite{ekera2017,ekera2020} steadily reduced factoring resource estimates. These improvements culminated in the full-width modular multiplication design of Gidney and Eker{\aa}~\cite{ge2021}, which requires roughly $6{,}200$ logical qubits (20 million noisy physical qubits) and $2.7\times 10^9$ Toffoli gates. To break the $2n$--$3n$ logical-qubit barrier for an $n$-bit modulus, Chevignard, Fouque, and Schrottenloher~\cite{cfs2024} introduced approximate modular exponentiation in a residue number system (RNS), replacing wide modular multiplications with arithmetic modulo thousands of small primes and compressing logical space to $n/2+o(n)$ qubits at the price of substantially higher gate counts. Building on this RNS framework, Gidney~\cite{gidney2025} combined deferred measurement-based uncomputation~\cite{gidney2018halving,babbush2018,berry2019} and coset-masked accumulators~\cite{gidney2019} with magic-state cultivation~\cite{gidney2024cultivation} and yoked surface codes~\cite{gidney2025yoked} to bring RSA-2048 below one million physical qubits. In that serial RNS architecture, however, each of the roughly twenty thousand primes performs its own table lookup and pays repeated arithmetic conversion costs across stages, leaving logical gate counts at $4.5$--$11\times 10^9$ Toffolis ($6.4\times 10^9$ at Gidney's $1{,}435$-qubit operating point), well above full-width multiplication~\cite{ge2021}.

\begin{figure*}[t]
\centering
\includegraphics[width=\textwidth]{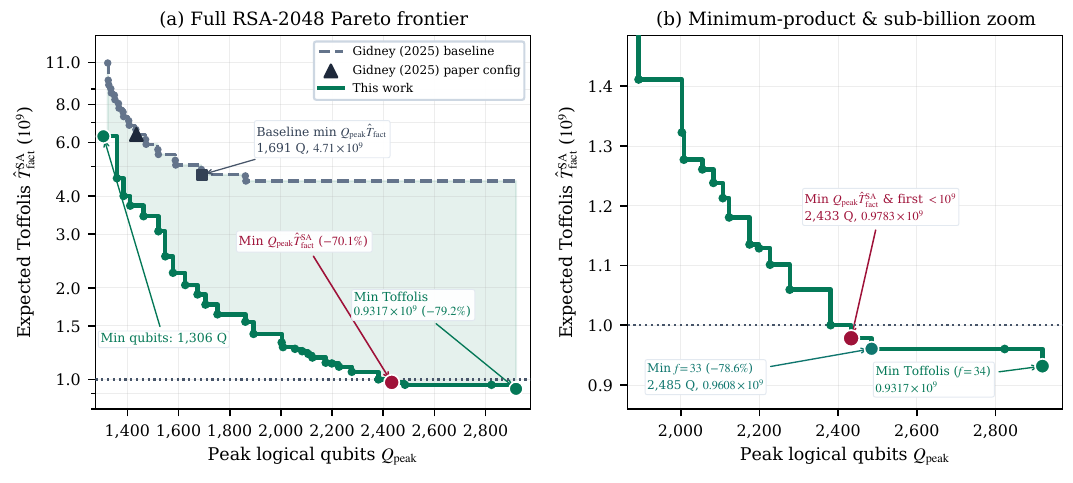}
\caption{Pareto frontier of compiled RSA-2048 modular-exponentiation circuits under a batch cap of $k\le 24$, plotting expected factoring Toffolis $\Tf^{\mathrm{SA}}$ (under Gidney's matched $b-2$ adder convention) against peak logical qubits $\Qp$ ($29$ Pareto-optimal points under $\Tf^{\mathrm{SA}}$ out of \CmbNumPareto{} under strict bit-level counting). (a)~Full log-scale comparison against the \NumRefFrontier-point Pareto frontier from our clean-ancilla sweep of \NumRefValid{} Gidney~\cite{gidney2025} baseline circuits (gray dashed); the dark triangle marks Gidney's simulated $(s=8,\ell=21,w=6)$ point at $1{,}435$ qubits (versus $1{,}399$ from Gidney's closed-form formula, which omits $36$ temporary clean ancillas). (b)~Linear zoom over the minimum-product and sub-billion regime, highlighting the $2{,}433$- and $2{,}485$-qubit milestones (accumulator width $f=33$) that we certify across all primes, together with the $2{,}919$-qubit endpoint at $f=34$. Counts $\Tf$ taken directly from emitted bit-level gate streams are $1.0\%$--$2.4\%$ higher and are compared in Sec.~\ref{sec:results}, Appendix~\ref{app:tables}, and Appendix~\ref{app:external}.}
\label{fig:pareto}
\end{figure*}

Recently, program search guided by large language models (LLMs) and evolutionary algorithms has uncovered new mathematical constructions~\cite{romera2024,novikov2025} and compact quantum circuits~\cite{gu2026}. In particular, Gu \emph{et al.}~\cite{gu2026} showed that automated search can improve standalone quantum subroutines, such as prime-field inversion, syndrome decoders, elliptic-curve point addition, and fixed-point arithmetic, while Khattar \emph{et al.}~\cite{khattar2026} formalized a classically verifiable compilation framework and benchmark suite in which candidate edits are checked by simulating sampled trajectories with tracked measurement phases. Extending automated discovery from isolated subroutines to a full RSA-2048 circuit introduces three obstacles that do not arise at component scale. First, a single run spans more than a thousand logical qubits and executes hundreds of millions of Toffolis across roughly twenty thousand primes. Because the four arithmetic stages share temporary ancilla qubits and rely on measurement-based uncomputation~\cite{jones2013,gidney2018halving,babbush2018,berry2019,jeon2026,yang2026}, local edits to one stage often fail to lower the peak qubit count or leave uncanceled phases downstream. Second, narrowing intermediate sum registers saves qubits across the circuit, but introduces rare arithmetic overflows near probability $10^{-10}$ that random simulation cannot reliably detect. Third, checking every candidate edit across all twenty thousand primes during search is prohibitively slow.

Here we address both the gate-count penalty of RNS factoring~\cite{cfs2024,gidney2025} and the scaling barriers of automated quantum circuit discovery~\cite{gu2026,khattar2026}. For verification, we adopt the binary wire-stream format of Gu \emph{et al.}~\cite{gu2026} and extend the phase-tracked trajectory simulation of Khattar \emph{et al.}~\cite{khattar2026} with exact symbolic tracking over all $2^M$ mid-circuit measurement outcomes, stratified screening during search, and analytical tail bounds for compressed registers. Guided by these checks, LLM-driven search reorganizes the entire four-stage RNS pipeline rather than tuning isolated subroutines. The largest gate reduction comes from sharing each exponent-window lookup across batches of up to $k=24$ primes instead of processing primes serially as in Ref.~\cite{gidney2025}. Sustaining large batches without excessive ancilla growth or uncertified wraparound errors relies on three coupled changes across the pipeline: aligning numerical representations between adjacent stages, localizing measurement-based uncomputation inside each window and prime, and bounding rare overflow probabilities analytically.

Implementing these mechanisms in a unified deterministic compiler produces a Pareto frontier of RSA-2048 circuits across $\CmbMinQ$--$\CmbMaxQ$ logical qubits under $k\le 24$ (Fig.~\ref{fig:pareto}; Sec.~\ref{sec:results}). Under identical clean-ancilla rules and Gidney's matched standard-adder convention $\Tf^{\mathrm{SA}}$ ($b-2$ Toffolis per $b$-bit adder), our frontier pushes below Gidney's~\cite{gidney2025} clean-ancilla minimum ($\CmbMinQ$ versus $1{,}323$ qubits), cuts the minimum qubit--Toffoli product $\Qp\Tf^{\mathrm{SA}}$ by $70.1\%$, and reduces expected factoring Toffolis by up to $78.6\%$ at matched accumulator precision, or $79.2\%$ with one additional accumulator qubit. This brings RSA-2048 factoring below one billion Toffolis for the first time, improving simultaneously in both qubits and gate count over full-width modular multiplication~\cite{ge2021}; across Gidney's qubit range, our frontier also strictly dominates every serial RNS baseline in Ref.~\cite{gidney2025}.

\section{Cross-Stage Design Principles}
\label{sec:principles}

In approximate RNS modular exponentiation for a 2048-bit modulus $N$, factoring requires $s+1$ valid Eker{\aa}--H{\aa}stad runs over $W$ exponent windows of width $w$ bits. Each run processes tens of thousands of $\ell=20$-bit primes $p$ through four stages: (i)~batched discrete-logarithm lookup and summation into $k$ register lanes, (ii)~in-place exponent reduction modulo $p-1$, (iii)~windowed exponentiation modulo $p$, and (iv)~fused Chinese Remainder Theorem (CRT) accumulation into an $f$-qubit coset-masked accumulator, followed by measurement-based cleanup and reverse Stage-(ii) uncomputation. Appendix~\ref{app:cost} details the circuit topology, qubit accounting, gate conventions, and error budget, while Table~\ref{tab:arch_comp} compares our architecture with Gidney~\cite{gidney2025} and Sharma and Lau~\cite{sharma2026}. Sharing each Stage-(i) quantum read-only memory (QROM)~\cite{babbush2018} lookup across a batch of $k$ primes (also explored concurrently up to $k=8$ in Ref.~\cite{sharma2026}) divides its per-prime Toffoli cost by $k$. Below, we describe the three cross-stage mechanisms that make large batches practical without excessive ancilla growth or uncertified errors (Fig.~\ref{fig:mech}).

\begin{table*}[tp]
\centering
\caption{Architectural comparison of RNS approximate modular exponentiation for RSA-2048 across Gidney~\cite{gidney2025}, concurrent work by Sharma and Lau~\cite{sharma2026}, and this work (matched $b-2$ and strict bit-level counts in the bottom row).}
\label{tab:arch_comp}
\footnotesize
\begingroup
\setlength{\tabcolsep}{4.2pt}
\renewcommand{\arraystretch}{1.14}
\newcommand{\tc}[2]{\parbox[t]{#1}{\raggedright\strut #2\strut}}
\begin{tabular}{@{}llll@{}}
\toprule
\tc{0.165\textwidth}{\textbf{Architectural dimension}} &
\tc{0.215\textwidth}{\textbf{Gidney (2025)~\cite{gidney2025}}} &
\tc{0.220\textwidth}{\textbf{Sharma \& Lau (2026)~\cite{sharma2026}}} &
\tc{0.340\textwidth}{\textbf{This work}} \\
\midrule
\tc{0.165\textwidth}{\textbf{Batch sharing $k$}} &
\tc{0.215\textwidth}{$k=1$ (serial per prime)} &
\tc{0.220\textwidth}{$k\in\{1,\dots,8\}$ ($s=8$ fixed)} &
\tc{0.340\textwidth}{$k\in\{1,\dots,24\}$ (rolling/bilinear ancillas independent of $k$)} \\
\tc{0.165\textwidth}{\textbf{Stage-(i) QROM \& lanes}} &
\tc{0.215\textwidth}{Serial QROM; uncompressed $\approx 32$-bit lane, prime deltas} &
\tc{0.220\textwidth}{Shared QROM, prime deltas, packed $k$-lane buffer ($29$--$30$ bits, no tail compression)} &
\tc{0.340\textwidth}{Compressed $D\in\{26,27\}$ lanes with certified tails; rolling ANF, bilinear ($w{=}6$), zero-tail packed ($w{\in}\{8,9\}$)} \\
\tc{0.165\textwidth}{\textbf{Deferred phase $\mathcal{K}_{\mathrm{glob}}$}} &
\tc{0.215\textwidth}{Deferred global $\mathcal{K}_{\mathrm{glob}}$ at run exit} &
\tc{0.220\textwidth}{Deferred global $\mathcal{K}_{\mathrm{glob}}$ at run exit} &
\tc{0.340\textwidth}{Eliminated ($0$ seg.\ cost) for rolling/bilinear; compact final sweep for packed} \\
\tc{0.165\textwidth}{\textbf{Stage-(ii) $\bmod\,(p{-}1)$ div.}} &
\tc{0.215\textwidth}{Restoring binary division} &
\tc{0.220\textwidth}{Trimmed restoring division} &
\tc{0.340\textwidth}{Offset unsigned lane + non-restoring $\mathrm{ctz}(p{-}1)$ odd-add ($238$ vs.\ $504$ Toff./prime, $+0\text{ Q}$)} \\
\tc{0.165\textwidth}{\textbf{Stage-(iii) $\bmod\,p$ exp.}} &
\tc{0.215\textwidth}{Canonical $\bmod\,p$ per step (deferred MBU cleanup)} &
\tc{0.220\textwidth}{Lazy $\bmod\,p$ + checkpoints} &
\tc{0.340\textwidth}{Shared-selector SelectSwap + 22--23-bit non-canon.\ + bilinear reverse cleanup} \\
\tc{0.165\textwidth}{\textbf{Stage-(iv) CRT accumulator}} &
\tc{0.215\textwidth}{Unfused 5-bit $\bmod\,N_{\mathrm{trunc}}$ (1 borrow Q)} &
\tc{0.220\textwidth}{Unfused $\bmod\,N_{\mathrm{trunc}}$ (1 borrow Q)} &
\tc{0.340\textwidth}{Fused 3-bit dyadic torus $\bmod\,2^f$ ($653$ at $b_0{=}9$, $0$ borrow, $\tfrac{2}{3}\times$ per-slice diam.\ at $g_{\mathrm{gap}}{=}f$)} \\
\tc{0.165\textwidth}{\textbf{Phase synth.\ \& verification}} &
\tc{0.215\textwidth}{$\lfloor\sqrt{K}\rfloor$ macro estimate} &
\tc{0.220\textwidth}{Macro estimate (local explicit-gate audit)} &
\tc{0.340\textwidth}{Explicit Half-DFS + bilinear cleanup; 6 all-prime 32-traj.\ + 3 batch-0 dual-verified} \\
\tc{0.165\textwidth}{\textbf{Best RSA-2048}\\\textbf{$\Qp$, $\Tf$, $\Qp\Tf$}} &
\tc{0.215\textwidth}{$1{,}691\text{ Q}$, $4.71\times 10^9$, $7.96\times 10^{12}$ (min $4.49\times 10^9$)} &
\tc{0.220\textwidth}{$1{,}806\text{ Q}$, $2.55\times 10^9$, $4.60\times 10^{12}$} &
\tc{0.340\textwidth}{$\mathbf{2{,}433\text{ Q}}$, $\mathbf{0.978\times 10^9}$, $\mathbf{2.380\times 10^{12}}$ ($b{-}2$; min $0.9608\times 10^9$ at $f{=}33$, $0.9317\times 10^9$ at $f{=}34$; strict $1.001\times 10^9$, min $0.9496\times 10^9$)} \\
\bottomrule
\end{tabular}
\endgroup
\end{table*}

\subsection{Aligning representations across stages}
\label{sec:align_repr}

\begin{figure*}[t]
\centering
\includegraphics[width=\textwidth]{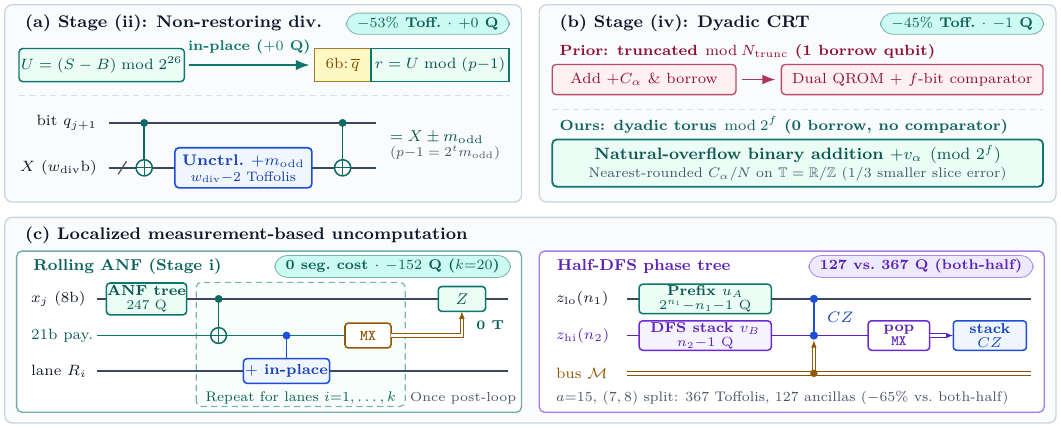}
\caption{Core cross-stage mechanisms from Secs.~\ref{sec:align_repr} and~\ref{sec:localize_uncomp}. (a)~Offset unsigned lane ($26$-bit, $\ell=20$) paired with non-restoring division that exploits $\mathrm{ctz}(p-1)$, replacing restoring adders with a single uncontrolled $+m_{\mathrm{odd}}$ adder per step. (b)~Dyadic CRT accumulation modulo $2^f$ on the torus using natural overflow and nearest rounding, which removes the borrow qubit and comparator while cutting the per-slice error diameter by $1/3$. (c)~Window cycle for rolling ANF in Stage~(i) with immediate Clifford \gate{Z} cleanup (left), and a configurable $(n_1,n_2)$ Half-DFS phase tree for Stage~(iii) (right).}
\label{fig:mech}
\end{figure*}

A recurring source of overhead in RNS factoring circuits is a mismatch between the numerical representation natural to one stage and that required by the next. Centering each discrete-logarithm sum $S$ around its mean $\mu=\lfloor\mathbb{E}[S]\rceil$ allows us to store it in $D\in\{26,27\}$ bits instead of the $32$-bit uncompressed lane of Ref.~\cite{gidney2025}. If this centered sum is stored as a signed two's-complement integer, however, Stage~(ii) cannot perform unsigned modulo-$(p-1)$ reduction in place without allocating an extra copy of the register. Instead, we choose a divisor-aligned half-width $H$ around $\mu$, shift $S$ by $B = \mu - H$, and store the result as a $D$-bit unsigned integer (Fig.~\ref{fig:mech}(a)):
\begin{equation}
\begin{aligned}
H &= 2^{D-\ell-1}(p-1),\quad B = \mu - H, \\
U &= (S - B)\bmod 2^D.
\end{aligned}
\label{eq:unsigned}
\end{equation}
For any nonoverflowing sum $S\in[\mu-H,\mu+H)$ within half-width $H$ of the mean, the shifted integer $U = S-B$ satisfies $0\le U<2H<2^D$, allowing modulo-$(p-1)$ reduction to run strictly in place on the $D$ lane wires. Using a primitive root $g$ modulo $p$, the constant offset $B$ folds classically into the $b_0$-bit Stage-(iii) startup lookup table $g^{r_{0\dots b_0-1}+B}\bmod p$, where $r_{0\dots b_0-1}$ denotes the lowest $b_0$ bits of the reduced exponent $r=U\bmod(p-1)$, leaving only one unconditional $D$-bit constant shift per prime (Proposition~1, Appendix~\ref{app:prop1}).

Even with an unsigned lane, standard restoring trial subtraction in Stage~(ii) requires two adders per division step, totaling $504$ Toffolis per prime (forward plus reverse) at $D=26$ bits. Building on quantum non-restoring division~\cite{thapliyal2021}, we factor $p-1 = 2^t m_{\mathrm{odd}}$ classically, where $m_{\mathrm{odd}}$ is odd and $t=\mathrm{ctz}(p-1)\ge 1$ counts the trailing binary zeros of $p-1$. Because those trailing zeros leave the lowest $j+t$ wires untouched at step $j$, each division step acts only on a narrowed slice $X$ of width $w_{\mathrm{div}} = \ell+1-t$ bits and conjugates a single \emph{uncontrolled} adder for the odd constant $+m_{\mathrm{odd}}$ by transversal \gate{CX} gates from the previous quotient bit $q_{j+1}$:
\begin{equation}
\begin{aligned}
&\Bigl(\bigl((X \oplus q_{j+1}^{\otimes w_{\mathrm{div}}}) + m_{\mathrm{odd}}\bigr)\bmod 2^{w_{\mathrm{div}}}\Bigr) \oplus q_{j+1}^{\otimes w_{\mathrm{div}}} \\
&\quad \equiv
\begin{cases}
X + m_{\mathrm{odd}} \pmod{2^{w_{\mathrm{div}}}}, & q_{j+1}=0,\\
X - m_{\mathrm{odd}} \pmod{2^{w_{\mathrm{div}}}}, & q_{j+1}=1,
\end{cases}
\end{aligned}
\label{eq:nonrestoring}
\end{equation}
followed by one final controlled restoration (Fig.~\ref{fig:mech}(a); full step-by-step construction and bijection proof on all of $\{0,1\}^D$, including wrapped states $U\ge 2H$, are given in Proposition~2, Appendix~\ref{app:prop2_loop2}). Across all $19{,}878$ primes of our primary $s=4,w=8$ family, this cuts forward-plus-reverse Stage~(ii) by more than half, from $504$ to an average of $238$ Toffolis per prime ($5.29\times 10^6$ Toffolis saved per run).

A second representation mismatch occurs at Stage~(iv). Accumulating CRT slices modulo the truncated modulus $N_{\mathrm{trunc}}=\lfloor N/2^{2048-f}\rfloor$ requires a persistent borrow qubit, a per-slice dual lookup, and an $f$-bit comparator, totaling $1{,}183$ emitted Toffolis per prime across eight fused slices, while incurring wraparound drift and floor bias up to $2\cdot 2^{-f}$ per slice (Fig.~\ref{fig:mech}(b)). Instead, we map each lifted CRT slice $C_\alpha/N$ ($C_\alpha\in[0,N)$) directly to its nearest $f$-bit dyadic grid point $v_\alpha/2^f$ on the torus $\mathbb{T}=\mathbb{R}/\mathbb{Z}$:
\begin{equation}
v_\alpha := \left\lfloor \frac{2^f C_\alpha + \lfloor N/2\rfloor}{N}\right\rfloor \bmod 2^f \in \{0,\dots,2^f-1\},
\label{eq:dyadic_vm}
\end{equation}
and accumulate $\sum_{\alpha} v_\alpha \pmod{2^f}$ using binary addition with natural overflow, requiring zero borrow qubits and no comparator. Integer multiples of $N$ vanish modulo $1$ on $\mathbb{T}$, eliminating wraparound drift, while symmetric rounding restricts each slice error to $[-\frac{1}{2}2^{-f},+\frac{1}{2}2^{-f}]$ (Proposition~3, Appendix~\ref{app:prop3_dyadic}). This cuts the worst-case per-slice error diameter by one-third at equal slice count and RNS guard precision $g_{\mathrm{gap}}=f$ (Eq.~\eqref{eq:tfact_dyadic}, Appendix~\ref{app:baseline_sweep}), and lowers Stage~(iv) from $1{,}183$ to $653$ emitted Toffolis per prime at $f=33,b_0=9$ ($442$ at $b_0=10$).

\subsection{Localizing measurement-based uncomputation}
\label{sec:localize_uncomp}

Even with aligned arithmetic, batching $k$ primes together still faces two workspace and cleanup bottlenecks. First, a packed QROM buffer requires $21k$ output wires, growing linearly with the prime batch size $k$. Second, deferring Stage-(i) measurement phase cleanup to the end of the run forces costly global phase sweeps and verification boundary flushes. We eliminate both bottlenecks by localizing lookup uncomputation and phase synthesis inside each window and prime.

For large batches at $w=8$, we replace the $k$-lane packed buffer with a rolling algebraic-normal-form (ANF) lookup (Fig.~\ref{fig:mech}(c)). For each exponent window, the circuit computes the Boolean products of the window bits once and reuses a single output register to stream the lookup values sequentially across all $k$ lanes. Each lane executes a Clifford \gate{CX} load, in-place addition, and $X$-basis measurement (\gate{MX}). Once all lanes are processed, a single sweep of Clifford \gate{Z} gates conditioned on the measurement outcomes cleans up the live subset wires within the same window. Including the global phase flip from the constant (empty-subset) term\footnote{An outcome-dependent global phase is unobservable in an uncontrolled run, but tracking it guarantees strict unitary phase cancellation across all $2^M$ mid-circuit measurement outcomes and preserves validity under coherent control.} removes deferred Stage-(i) phase cleanup entirely, so the Toffoli count under segmented verification matches monolithic execution, $T_{\mathrm{seg}}=T_{\mathrm{mono}}$ (Proposition~4(b), Appendix~\ref{app:prop4_phase}). This decouples Stage-(i) scratch from $k$, saving $152$ qubits at $k=20$ and letting $k\le 24$ fit below $2{,}500$ qubits at $s=4,w=8$. At the extremes of the frontier, low-workspace schedules use a bilinear ANF split with zero segmentation overhead, while wide $w=9$ schedules pair naturally with compact packed buffers that omit unused trailing slots under our $k\le 24$ cap (Appendix~\ref{app:layout_proj}).

\begin{figure*}[t]
\centering
\includegraphics[width=\textwidth]{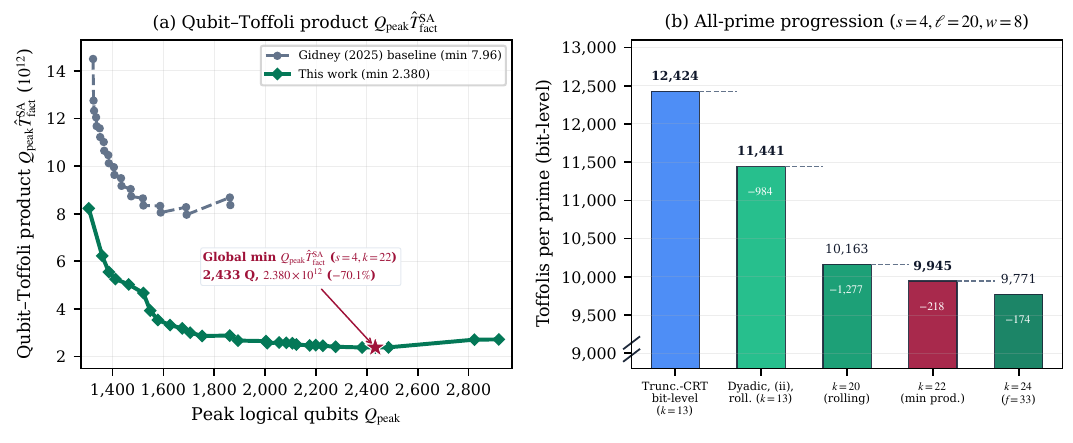}
\caption{Qubit--Toffoli product and controlled progression across all primes. (a)~Matched $b-2$ qubit--Toffoli product $\Qp\Tf^{\mathrm{SA}}$ across Gidney's~\cite{gidney2025} baseline and this work, reaching its global minimum of $2.380\times 10^{12}$ ($-70.1\%$) at $2{,}433$ qubits ($s=4,k=22$). (b)~Exact bit-level Toffoli count per prime averaged over all $19{,}878$ primes at $(s=4,\ell=20,w=8,f=33)$, moving from the mod-$N_{\mathrm{trunc}}$ baseline at $k=13$ to dyadic rolling ANF at $k=13,20,22,24$ (step deltas rounded from exact unrounded averages; truncated axis marked at bottom).}
\label{fig:ablations}
\end{figure*}

Two further primitives keep temporary workspace small across every schedule. Adding signed prime differences into wider lanes without sign-extension wires saves $5$--$6$ ancillas (Proposition~5, Appendix~\ref{app:prop5_trunc}). In Stage~(iii), measuring intermediate mod-$p$ words early frees live registers and merges their phase corrections into a single $a$-bit Boolean table. Splitting that table address into a materialized $n_1$-bit prefix tree and a depth-first $n_2$-bit suffix stack (Half-DFS; Fig.~\ref{fig:mech}(c) and Proposition~4(a), Appendix~\ref{app:prop4_phase}) cuts the scratch space for a $15$-bit phase table from $367$ to $127$ qubits relative to materializing both halves.

\subsection{Certifying rare overflow tails beyond sampling}
\label{sec:cert_tails}

Compressing each discrete-logarithm lane to $D=26$ or $27$ bits saves logical qubits across all $k$ lanes of each batch, but introduces rare wraparound events at probabilities around $10^{-10}$ or below that no tractable random trajectory sample can reliably detect. For $27$-bit lanes, classical Hoeffding bounds~\cite{hoeffding1963} already yield negligible single-run trace distance, such as $\eta\le 1.75\times 10^{-7}$ (multi-run bound $(s+1)\eta\le 7.0\times 10^{-7}$) at $s=3,w=9$. Reaching $26$-bit lanes exploits the fact that nearly uniform discrete logarithms have roughly one-third the variance of the worst-case endpoint distribution: because the uniform input state renders the window discrete logarithms mutually independent for each prime, we enumerate all $2^{w_j}$ table entries per prime and window ($w_j\le w$) at compile time to certify moment-generating-function (MGF) bounds with upward rounding, giving $\eta\le 3.37\times 10^{-5}$ on our primary $s=4,w=8$ family (Appendix~\ref{app:mgf}).

Because Propositions~1, 2, and~4 guarantee clean ancilla reset and input-independent phase cancellation even on wrapped branches, a Stinespring dilation argument yields the coherent channel trace-distance bound $2\sqrt{p_{\mathrm{bad}}}\le\eta$ in Eq.~\eqref{eq:app_trace_bound}, which enters the retention probability in Eq.~\eqref{eq:tfact_dyadic}. Although our concrete RSA-2048 gate-stream benchmarks evaluate the standard single-base proxy $3^x\bmod N$ following Ref.~\cite{gidney2025}, substituting any Eker{\aa}--H{\aa}stad two-base instance $g^{e_1}h^{-e_2}\bmod N$ leaves the circuit topology, peak logical qubit count $\Qp$, and Stage-(ii)--(iv) gate structure strictly invariant; only prime filtering and the instance-specific MGF tail certificate for $D=26$ need to be recomputed at compile time, which takes seconds, whereas $D=27$ Hoeffding bounds hold universally across all moduli and bases (Appendix~\ref{app:twobase}).

\section{Sub-Billion RSA-2048 Frontier and All-Prime Certification}
\label{sec:results}

\subsection{Controlled progression below one billion Toffolis}
\label{sec:case_study}

To isolate how the cross-stage mechanisms of Sec.~\ref{sec:principles} drive factoring below one billion Toffolis, Fig.~\ref{fig:ablations}(b) tracks the exact bit-level Toffoli count per prime across all $19{,}878$ primes of our primary $(s=4,\ell=20,w=8,f=33)$ family. Starting from our bit-level mod-$N_{\mathrm{trunc}}$ implementation at $k=13$ ($2{,}204$ qubits), the circuit costs $12{,}424$ emitted Toffolis per prime and incurs $3.23\times 10^6$ extra Toffolis per run to flush phase corrections at segment boundaries. At the same batch size $k=13$, adding offset unsigned lanes with non-restoring division, dyadic torus CRT, and rolling ANF lowers the per-prime cost by $7.9\%$ to $11{,}441$ Toffolis, eliminates all deferred Toffolis at segment boundaries, and saves $5$ logical qubits. Together with the higher retention probability of dyadic CRT relative to the fused 3-bit baseline modulo $N_{\mathrm{trunc}}$, this reduces expected factoring Toffolis by $8.1\%$ before increasing $k$.

\begin{table*}[t]
\caption{Nine milestone circuits from our \CmbNumPareto-point RSA-2048 Pareto frontier compiled at the bit level under $k\le 24$ ($29$ points under $\Tf^{\mathrm{SA}}$; complete listing in Appendix~\ref{app:tables}), contrasted against four reference points from our clean-ancilla sweep of Gidney~\cite{gidney2025}. Columns report matched $b-2$ counts $\Tf^{\mathrm{SA}}$ and $\Qp\Tf^{\mathrm{SA}}$, the Toffoli reduction $\Delta\widehat{T}_{\mathrm{ref}}^{\mathrm{SA}}$ at equal or fewer qubits, and strict bit-level counts $\Tf$. All compiled rows use $\ell=20$-bit primes and offset unsigned lanes ($D\mathrm{u}$), with $f=33$ except the last two $w=9$ rows at $f=34$. $^{\mathrm{a}}$Uncompressed serial ($k{=}1$) Gidney baseline at $f=33$ evaluated in our clean-ancilla simulator (macro-modeled $\Tf^{\mathrm{SA}}$). $^{\mathrm{b}}$$1{,}306$ qubits lies $17$ qubits below Gidney's $1{,}323$-qubit clean-ancilla minimum ($-42.6\%$ below Gidney's $1{,}323$-qubit point). Stars $^\star$ mark circuits certified across all primes over $32$ trajectories (Sec.~\ref{sec:gate_level}), daggers $^\dagger$ mark milestones verified on batch~0 with both backends, and boldface marks the global minimum $\Qp\Tf^{\mathrm{SA}}$.}
\label{tab:frontier}
\footnotesize
\begin{tabular*}{\textwidth}{@{\extracolsep{\fill}}lcrrrrr@{}}
\toprule
Compiled circuit ($\star$ all-prime 32-traj.; $\dagger$ dual) & $(s,k,w,D,f)$ & $\Qp$ & $\Tf^{\mathrm{SA}}$ ($10^9$) & $\Qp\Tf^{\mathrm{SA}}$ ($10^{12}$) & $\Delta\widehat{T}_{\mathrm{ref}}^{\mathrm{SA}}$ & $\Tf$ ($10^9$) \\
\midrule
Gidney (2025) min-$\Qp$ baseline$^{\mathrm{a}}$ & $(14,1,7,\ell{=}19)$ & $1{,}323$ & $10.9578$ & $14.497$ & \textemdash & \textemdash \\
Gidney (2025) published config$^{\mathrm{a}}$ & $(8,1,6,\ell{=}21)$ & $1{,}435$ & $6.3879$ & $9.167$ & \textemdash & \textemdash \\
Gidney (2025) min-$\Qp\Tf$ baseline$^{\mathrm{a}}$ & $(4,1,6,\ell{=}21)$ & $1{,}691$ & $4.7053$ & $7.957$ & \textemdash & \textemdash \\
Gidney (2025) min-$\Tf$ baseline$^{\mathrm{a}}$ & $(3,1,6,\ell{=}21)$ & $1{,}863$ & $4.4869$ & $8.359$ & \textemdash & \textemdash \\
\midrule
Bilinear low-workspace $k{=}1$ (min-$\Qp$)$^{\star}$ & $(14,1,6,26\mathrm{u},33)$ & $1{,}306$ & $6.2943$ & $8.220$ & \textemdash$^{\mathrm{b}}$ & $6.3548$ \\
Low-workspace rolling $k{=}2$ ($w{=}6$)$^{\dagger}$ & $(14,2,6,26\mathrm{u},33)$ & $1{,}359$ & $4.5828$ & $6.228$ & $-44.8\%$ & $4.6432$ \\
Zero-tail packed $k{=}4$ ($s{=}8, w{=}8$) & $(8,4,8,27\mathrm{u},33)$ & $1{,}548$ & $2.5385$ & $3.930$ & $-53.7\%$ & $2.5679$ \\
Rolling ANF $k{=}13$, $0$ seg.\ cost ($s{=}4$)$^{\star}$ & $(4,13,8,26\mathrm{u},33)$ & $2{,}199$ & $1.1289$ & $2.482$ & $-74.8\%$ & $1.1517$ \\
Rolling ANF $k{=}20$, $0$ seg.\ cost ($s{=}4$)$^{\star}$ & $(4,20,8,26\mathrm{u},33)$ & $2{,}381$ & $1.0003$ & $2.382$ & $-77.7\%$ & $1.0232$ \\
Rolling ANF $k{=}22$ (min-$\Qp\Tf^{\mathrm{SA}}$, first $<10^9$)$^{\star}$ & $(4,22,8,26\mathrm{u},33)$ & $\mathbf{2{,}433}$ & $\mathbf{0.9783}$ & $\mathbf{2.380}$ & $-78.2\%$ & $\mathbf{1.0012}$ \\
Rolling ANF $k{=}24$ (min $\Tf^{\mathrm{SA}}$ at $f{=}33$, first strict $<10^9$)$^{\star}$ & $(4,24,8,26\mathrm{u},33)$ & $2{,}485$ & $0.9608$ & $2.388$ & $-78.6\%$ & $0.9836$ \\
Wide-window $w{=}9, k{=}22$ ($f{=}34$)$^{\star}$ & $(3,22,9,27\mathrm{u},34)$ & $2{,}823$ & $0.9604$ & $2.711$ & $-78.6\%$ & $0.9782$ \\
Wide-window $w{=}9, k{=}24$ (min-$\Tf^{\mathrm{SA}}$, $f{=}34$)$^{\dagger}$ & $(3,24,9,27\mathrm{u},34)$ & $2{,}919$ & $0.9317$ & $2.720$ & $-79.2\%$ & $0.9496$ \\
\bottomrule
\end{tabular*}
\end{table*}

Crucially, because rolling ANF decouples Stage-(i) scratch from $k$ while MGF certification holds every lane to $26$ bits, scaling batch sharing up to $k=24$ then drives the per-prime cost monotonically down to $9{,}771$ Toffolis, crossing below one billion expected factoring Toffolis at $2{,}433$ qubits under matched $b-2$ accounting and at $2{,}485$ qubits under strict bit-level counting. Every mechanism is indispensable: without rolling ANF and lane compression, $k=24$ would exceed $2{,}500$ qubits, and without non-restoring division and dyadic CRT, the per-prime kernel would keep expected factoring Toffolis above one billion even at $k=24$.

\subsection{Compiled Pareto frontier and external comparisons}
\label{sec:pareto_results}

Extending this compilation across $(s,k,w,D,f)$ under $k\le 24$, compiling \CmbNumConfigs{} configurations at the bit level yields \CmbNumPareto{} Pareto-optimal circuits across $\CmbMinQ$--$\CmbMaxQ$ logical qubits (\CmbNumParetoEqBudget{} at Gidney's fixed 33-bit accumulator budget; $29$ remain Pareto-optimal under matched $b-2$ accounting $\Tf^{\mathrm{SA}}$ in Figs.~\ref{fig:pareto} and~\ref{fig:ablations}(a), Table~\ref{tab:frontier}, and Appendix~\ref{app:tables}). Under matched $b-2$ accounting and identical clean-ancilla rules, our frontier improves on Gidney's \NumRefValid-circuit baseline across the entire qubit range and divides naturally into three regimes.

At the low-qubit extreme, bilinear ANF and 2-ancilla Cuccaro arithmetic reach $\CmbMinQ$ logical qubits ($17$ below Gidney's clean-ancilla minimum; Appendices~\ref{app:layout_proj}, \ref{app:baseline_sweep}, and~\ref{app:tables}). Enabling minimal batching ($k=2$) at $1{,}359$ qubits already cuts expected Toffolis by $28\%$ relative to Gidney's simulated reference point while saving $76$ qubits.

As window width and batch size grow, the qubit--Toffoli product $\Qp\Tf^{\mathrm{SA}}$ enters a deep valley (Fig.~\ref{fig:ablations}(a)), reaching its global minimum of $2.380\times 10^{12}$ ($-70.1\%$ below Gidney's $\RefMinProdVal$) at $\CmbMinQTQ$ qubits ($k=22$) while crossing below one billion Toffolis. This valley reflects a tradeoff between two competing effects: sharing each Stage-(i) lookup across $k$ primes reduces the lookup cost per prime roughly as $1/k$, whereas each additional lane requires $D$ extra work qubits. Once these $1/k$ lookup savings begin to level off near $k\approx 20$--$22$, adding further lanes raises $\Qp$ proportionally faster than it lowers $\Tf^{\mathrm{SA}}$, causing the product to turn upward. Extending batching to $k=24$ at $2{,}485$ qubits reaches our fixed-$f=33$ minimum of $0.9608\times 10^9$ expected factoring Toffolis ($-78.6\%$), certified across all primes, and widening to $w=9$ with one additional accumulator qubit ($f=34$) lowers it further to $0.9317\times 10^9$ ($-79.2\%$) at $\CmbMaxQ$ qubits (Table~\ref{tab:frontier}).

Accounting and external checks confirm that these gains do not rely on a narrow cost convention (Appendix~\ref{app:external}). Under our stricter bit-level count $\Tf$, which charges every logical \gate{AND}, \gate{CCX}, and \gate{CSWAP}, totals rise by only $1.0\%$--$2.4\%$ and remain below one billion Toffolis at both $f=33$ and $f=34$ (Table~\ref{tab:frontier}). Restoring Gidney's symmetric $0.99$ classical lattice factor likewise keeps every sub-billion milestone below $10^9$, and comparison against Gidney's closed-form analytic formulas preserves a $\sim 70\%$ product reduction even with freely retuned accumulator precision. Against the $13$ RSA-2048 circuits at fixed $s=8$ reported concurrently by Sharma and Lau~\cite{sharma2026}, our frontier saves $24.9\%$--$43.4\%$ expected Toffolis at equal or fewer qubits even under our stricter bit-level counting.

\subsection{Gate-stream certification across all primes}
\label{sec:gate_level}

Every compiled circuit lowers explicitly to a bit-level gate stream in the binary wire format of Gu \emph{et al.}~\cite{gu2026}. Inside the arithmetic core, each of the $M$ mid-circuit \gate{MX} measurements acts on a wire whose computational-basis value is a deterministic Boolean function of surviving basis registers, and its classical outcome controls only diagonal Clifford phase gates. As a result, distinct computational-basis branches never merge, and assigning one independent $\mathbb{F}_2$ variable to each measurement outcome captures all $2^M$ measurement histories exactly. Exploiting this structure, we certify six representative circuits spanning $\CmbMinQ$ to $2{,}823$ qubits across all $119{,}482$ constituent 20-bit prime evaluations (summed across the six circuits) without extrapolation (Table~\ref{tab:all_prime}). Each circuit is partitioned into clean batch segments, and every segment is checked independently by two backends: a symbolic tracker that verifies zero residual phase over all $2^M$ measurement bitstrings, and a randomized \emph{kickmix} replay~\cite{khattar2026} over $32$ computational-basis trajectories (Appendix~\ref{app:verif}).

These full-width runs establish three empirical guarantees. First, every certified dyadic accumulator output matches the single-base proxy $(3^x\bmod N)/N$ (Appendix~\ref{app:twobase}) within $8.3\times 10^{-8}$ on $\mathbb{T}$, more than two orders of magnitude inside Proposition~3's worst-case bound. Second, all five rolling and bilinear circuits, including both fixed-$f=33$ sub-billion milestones, satisfy $T_{\mathrm{seg}}-T_{\mathrm{mono}}=0$ to the gate, while the tagged boundary flushes on the packed $w=9$ circuit merge into a single run-exit phase sweep via cross-batch commutativity, as verified directly on scaled 330-bit multi-segment chains (Appendix~\ref{app:calibration}). Third, stratified full-width compilation over five representative batches recovers the peak qubit count with zero error and matches the exact Toffoli count across all primes within $0.10\%$. Together with dual-backend verification of batch~0 on three additional milestones ($1{,}359$, $2{,}822$, and $2{,}919$ qubits), this validates the remaining interior points of our \CmbNumPareto-circuit frontier (Table~\ref{tab:cmb_full}; Appendix~\ref{app:calibration}).

\section{Discussion and Outlook}
\label{sec:discussion}

In this work, we reorganized the approximate RNS pipeline for RSA-2048 across stages using LLM-driven search, certified the compressed registers with analytical tail bounds, and verified the compiled circuits at the elementary gate level. Rather than tuning isolated subroutines, our gains come from combining large-batch lookup sharing (up to $k=24$) with aligned representations across stages, localized measurement-based uncomputation, and analytical overflow bounds. Together, these mechanisms cut the minimum qubit--Toffoli product by $70.1\%$ relative to Gidney's 2025 baseline and bring RSA-2048 factoring below one billion expected Toffolis for the first time.

Mapping this logical frontier onto planar surface-code hardware introduces two practical tradeoffs (Table~\ref{tab:hw_ext}(a), Appendix~\ref{app:external}). First, wider QROM sharing saves Toffoli gates at the expense of denser two-qubit Clifford fanout (\gate{CX} and \gate{CZ}) and mid-circuit $X$-basis measurements, lifting the ratio of explicit two-qubit Cliffords to Toffolis from roughly thirteenfold at $1{,}306$ qubits to nearly ninetyfold at wide batching. Although lattice-surgery parity measurements can broadcast \gate{CX} fanout in parallel, the net physical spacetime cost depends on routing layout, and rolling ANF trades lower qubit and \gate{CX} counts for $k$-fold in-window serialization. Second, choosing an operating point balances logical qubits against classical postprocessing and serialization. Larger $s$ reduces logical qubits while tightening the $(s+2)$-dimensional lattice problem~\cite{ekera2020,gidney2025}, favoring intermediate $s\approx 6$--$8$ on qubit-limited machines. Conversely, on larger hardware where in-window serialization is acceptable, relaxing $k\le 24$ lowers expected Toffolis below $0.90\times 10^9$ at $k=42$ without moving the minimum product at $k=22$ (Table~\ref{tab:hw_ext}(b)).

Looking ahead, three directions naturally follow from this study. First, coupling our deterministic compiler directly to physical surface-code routing, lattice-surgery scheduling, and reaction-time models will enable end-to-end physical optimization of $(s,w,k,b_0)$. Second, the cross-stage ideas developed here transfer directly to other arithmetic-heavy fault-tolerant tasks, including elliptic-curve discrete logarithms, windowed QROM pipelines, and phase-estimation arithmetic. Third, our workflow provides a practical template for scaling LLM-driven circuit discovery beyond standalone subroutines to full fault-tolerant algorithms by combining automated search with analytical error bounds, symbolic measurement-phase checks, and gate-level replay.

\emph{Note added.} During manuscript preparation, we became aware of the concurrent independent preprint of Sharma and Lau~\cite{sharma2026} studying shared QROM lookups and lazy modular residues for RSA-2048 across $1{,}432$--$1{,}806$ qubits, and added architectural and quantitative comparisons in Table~\ref{tab:arch_comp}, Sec.~\ref{sec:pareto_results}, and Appendix~\ref{app:external}.

\emph{AI Usage Disclosure.} The main algorithmic and circuit-compilation improvements in this work were generated using GPT-6 Astra in 2026 within the LLM-driven verifiable compilation workflow of Appendix~\ref{app:verif}, and AI tools assisted in manuscript writing and editing. All mathematical proofs, circuit implementations, gate-stream verification runs, and final claims were independently checked by the authors, who take full responsibility for the accuracy and integrity of this manuscript.

\emph{Code and Data Availability.} The compiler, dual-backend verifiers (symbolic and \emph{kickmix}), archived elementary-gate replay tools, tail checkers, prime inputs, frontier ledgers, receipts ($3.14\text{ TB}$ of streams), and figure scripts are available at \url{https://github.com/ZhangTao1999/LLM-Driven-Verifiable-Quantum-Compilation-for-RSA}. The internal search harness and cluster driver are not redistributed, and the Apache-2.0 \texttt{andigu/uncompute} backend~\cite{gu2026} is cited only; all counts, checks, and figures reproduce from this repository.

\begin{acknowledgments}
We thank Yingfei Gu, Xiaotian Nie, Yiqi Yu, Juan Yao, and Hui Zhai for helpful discussions.
\end{acknowledgments}

\clearpage
\bibliography{references}

\onecolumngrid
\clearpage
\let\addcontentsline\realaddcontentsline
\tableofcontents

\begin{appendix}
\setlist[itemize]{leftmargin=1.6em,itemsep=2.5pt,topsep=4pt}
\setlist[enumerate]{leftmargin=1.8em,itemsep=2.5pt,topsep=4pt}

\section{RNS Pipeline, Cost Conventions, and Dyadic Error Budget}
\label{app:cost}

This appendix establishes the full compilation model underlying Secs.~\ref{sec:principles} and~\ref{sec:results}. Appendix~\ref{app:stages} fixes the Eker{\aa}--H{\aa}stad RNS parameters and prime-pool capacity constraints; Appendix~\ref{app:layout_proj} details our four-stage register layout (Fig.~\ref{fig:circuit_overview}) and stratified batch projection; Appendix~\ref{app:lookup_regimes} specifies clean-ancilla tracking, strict emitted bit-level gate counting, and the matched $b-2$ conversion; Appendix~\ref{app:twobase} proves structural invariance when transferring from the single-base benchmark proxy $3^x\bmod N$ to arbitrary two-base instances; and Appendix~\ref{app:baseline_sweep} derives the dyadic error budget and documents our 28-circuit sweep of Gidney's baseline.

\needspace{5\baselineskip}
\subsection{Eker{\aa}--H{\aa}stad reduction and RNS arithmetic}
\label{app:stages}

Under the Eker{\aa}--H{\aa}stad short discrete-logarithm reduction~\cite{ekera2017,ekera2020} and Gidney's resource-estimation model~\cite{gidney2025}, factoring a 2048-bit RSA modulus $N$ reduces to collecting $R=s+1$ valid quantum runs controlled by a tradeoff integer $s\in\{3,\dots,14\}$. Each run evaluates $a_{\mathrm{base}}^{e_1} b_{\mathrm{base}}^{e_2}\bmod N$ over a concatenated superposed exponent $\ket{x}$ of width
\begin{equation}
m_e = m_1 + m_2 = \left\lceil \frac{2048(s+1)}{2s}\right\rceil + \left\lceil \frac{2048}{2s}\right\rceil \approx 1024 + \frac{2048}{s},
\label{eq:me}
\end{equation}
partitioned into $W=\lceil m_e/w\rceil$ windows $x_j$ of width $w_j\le w\in\{6,7,8,9\}$ bits (with a shorter final window when $w\nmid m_e$). Outer state-preparation Hadamards and the terminal semiclassical inverse quantum Fourier transform (QFT)~\cite{griffiths1996,ross2016} add zero peak logical qubits and $<10^5$ Toffolis, and are omitted symmetrically from candidate and baseline counts as in Ref.~\cite{gidney2025}.

Following Chevignard, Fouque, and Schrottenloher~\cite{cfs2024} and Gidney~\cite{gidney2025}, classical precomputation evaluates each $2^{w_j}$-entry window table $F_j(x_j)\bmod N < N$ and selects $|\mathcal{P}|\approx 1.5\text{--}2.2\times 10^4$ distinct $\ell=20$-bit odd primes so that the unreduced product is represented uniquely in the RNS:
\[
X_{\mathrm{prod}} = \prod_{j=0}^{W-1} F_j(x_j) < N^W \le L = \prod_{p\in\mathcal{P}} p,
\qquad
S_p(x) = \sum_{j=0}^{W-1} d_{p,j}(x_j).
\]
Compile-time filtering excludes any prime dividing a table entry ($1.6\%$ at $w=6$ and $6\%$--$12\%$ at $w\in\{8,9\}$). Using a primitive root $g_p$ modulo $p$, multiplication modulo $p$ reduces to summing $W$ discrete logarithms $S_p(x)$ (written $S$ when $p$ is clear from context), and numerical CRT reconstruction folds the per-prime residues additively into an $f$-qubit accumulator storing the leading $f$ bits of $(X_{\mathrm{prod}}\bmod N)/N$.

\leadin{Prime-pool capacity at $w=6$ and $w=9$.}
In the low-qubit $w=6$ regime ($s\le 14$, $m_e\ge 1{,}172$, $W\ge 196$), summing $\log_2 p$ across all $20{,}390$ primes in $(2^{18},2^{19})$ gives $378{,}296.44\text{ bits}$, strictly below $\log_2(N^{196})\ge 401{,}212\text{ bits}$ for any 2048-bit modulus $N\ge 2^{2047}$. Because compile-time prime filtering only lowers available capacity further, $w=6$ requires $\ell=20$-bit primes, whereas widening to $w=7$ reduces $W$ to $168$ and permits Gidney's 19-bit $(14,19,7)$ baseline. At the wide-window endpoint $(s=3,\ell=20,w=9)$ ($m_e=1{,}708$, $W=190$), deterministic RNS search constructs $|\mathcal{P}|=19{,}682$ distinct 20-bit primes (an $11.3\%$ reduction from $22{,}190$ primes at $w=8$) satisfying $\prod_{p\in\mathcal{P}}p\ge N^{190}$ and effective wraparound gap $\log_2(N/|L\bmod^{\pm}N|)=40.881\text{ bits}>34\text{ bits}$.

\needspace{5\baselineskip}
\subsection{Register layout and stratified batch projection}
\label{app:layout_proj}

\begin{figure}[!htbp]
\centering
\includegraphics[width=0.90\textwidth]{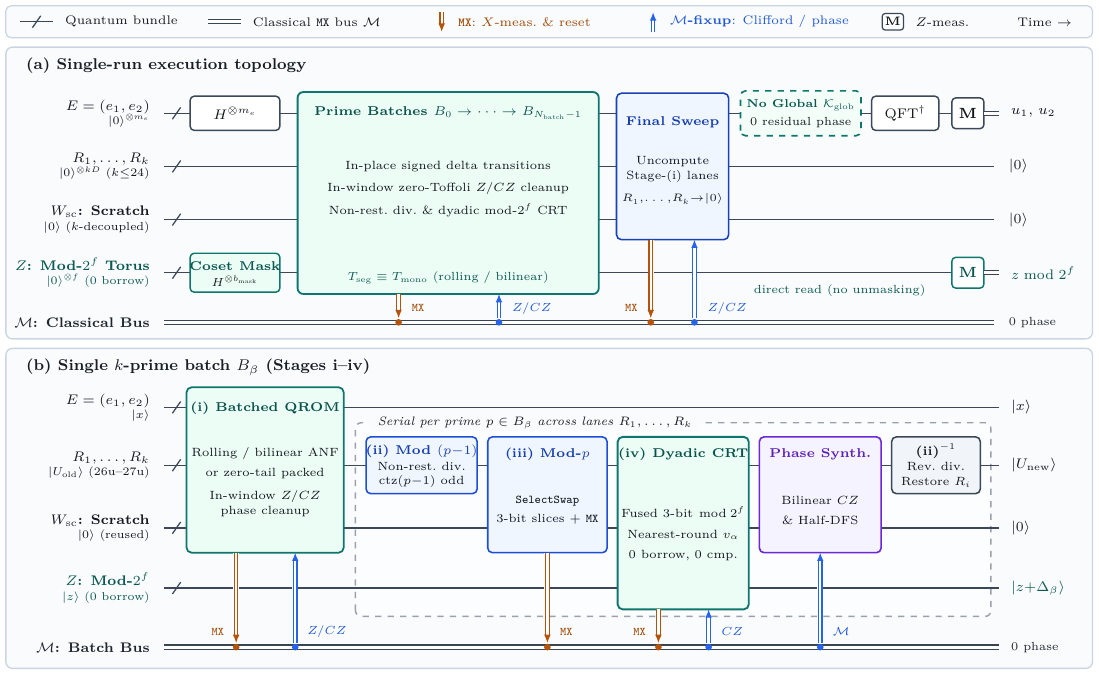}
\caption{Quantum circuit topology of our RSA-2048 compilation. (a)~Single-run layout showing in-place signed delta transitions across prime batches $B_0\to\cdots\to B_{N_{\mathrm{batch}}-1}$, final lane uncomputation, and terminal inverse QFT. Under rolling and bilinear ANF, immediate in-window Clifford cleanup eliminates the deferred global phase block $\mathcal{K}_{\mathrm{glob}}$ ($T_{\mathrm{seg}}\equiv T_{\mathrm{mono}}$), while mod-$2^f$ accumulation with natural overflow removes the borrow qubit and comparator. (b)~Single $k$-prime batch $B_\beta$ mapping Stages~(i)--(iv) with \gate{MX} ($\Downarrow$) and $\mathcal{M}$-conditioned Clifford ($\Uparrow$) feedback.}
\label{fig:circuit_overview}
\end{figure}

Figure~\ref{fig:circuit_overview} summarizes the quantum register layout across the exponent register $E=(e_1,e_2)$, $k$ compressed offset unsigned discrete-logarithm lanes $R_1,\dots,R_k$ of width $D$ (denoted $D\mathrm{u}$, with $D\in\{26,27\}$; $k\le 24$ on the main frontier), shared scratch register $W_{\mathrm{sc}}$, dyadic torus accumulator $Z$, and classical \gate{MX} bus $\mathcal{M}$. Within each $k$-prime batch $B_\beta$, the four stages operate as follows:
\begin{itemize}
\item \emph{Stage~(i) (Batched QROM and delta updates).} Lanes $R_1,\dots,R_k$ update in place via signed prime-to-prime differences across batches until one final uncomputation sweep. At $w=8$ ($k\ge 12$), rolling ANF materializes $2^8-8-1=247$ subset-product wires and streams a single 21-bit payload across all $k$ lanes ($292$ scratch qubits vs.\ $444$ packed at $k=20$). At $w=9$ under $k\le 24$, narrower zero-tail packed buffers omit unused trailing prime slots in the final partial batch ($462$--$504$ vs.\ $523$ qubits for a 9-bit rolling tree).
\item \emph{Stage~(ii) (In-place $\bmod\,(p-1)$ reduction).} Each offset unsigned lane reduces in place modulo $p-1$ via non-restoring $\mathrm{ctz}(p-1)$-sliced conjugated division (Proposition~2, Appendix~\ref{app:prop2_loop2}).
\item \emph{Stage~(iii) (Modular exponentiation $g^S\bmod p$).} A 3-bit selector ANF tree is shared across all source slices of SelectSwap quantum read-only access memory (QROAM)~\cite{berry2019,low2024} ($b_0\in\{8,9,10\}$). Intermediate sums remain noncanonical within $[0,8(p-1)]\subset[0,2^{23})$ with adders truncated to exact partial-sum widths; reverse cleanup uses bilinear Clifford gates, and low-workspace $w=6$ schedules use 2-ancilla Cuccaro adders~\cite{cuccaro2004}.
\item \emph{Stage~(iv) (Fused dyadic CRT and cleanup).} The final windowed multiplication fuses into $M_{\mathrm{CRT}}=8|\mathcal{P}|$ dyadic additions modulo $2^f$ on 3-bit slices, followed by Half-DFS phase cleanup and reverse Stage~(ii). Across all $19{,}878$ primes at $s=4,w=8,b_0=9$, the complete Stage-(ii)--(iv) per-prime kernel averages $2{,}963.0$ Toffolis per prime ($2{,}959.0$ at $b_0=10$).
\end{itemize}

\leadin{Disjoint gate partitioning and stratified projection.}
To prevent double-counting across nested calls, all emitted non-Clifford gates partition into four disjoint groups: (1)~batched Stage~(i) (batch~0 startup, interior $k$-prime transitions, and the final batch); (2)~the per-prime kernel (forward Stage~(ii), Stage~(iii), fused Stage-(iv) dyadic CRT, reverse cleanup, Half-DFS prefix phase synthesis, and reverse Stage~(ii)); (3)~final-batch lane uncomputation ($T_{\mathrm{I,fin}}$); and (4)~deferred global phase cleanup ($T_{\mathrm{glob}}=0$ under rolling and bilinear ANF).

Letting $N_{\mathrm{batch}}=\lceil|\mathcal{P}|/k\rceil$, stratified projection compiles batch~0 (giving $T_{\mathrm{I},0}, T_{\mathrm{pk},0}$), $n_{\mathrm{int}}$ evenly spaced interior batches spanning $p_0$ to $p_{\max}$ (giving per-batch mean $\overline{T}_{\mathrm{I,int}}$ and per-prime mean $\overline{T}_{\mathrm{pk,int}}$ averaged over $\mathrm{ctz}(p-1)$), and the final batch of $|\mathcal{P}_{\mathrm{last}}|$ primes (giving $T_{\mathrm{I,last}}, T_{\mathrm{pk,last}}$) to project unsegmented single-run Toffolis via
\begin{equation}
\begin{aligned}
\Ts &= T_{\mathrm{I},0} + (N_{\mathrm{batch}}-2)\,\overline{T}_{\mathrm{I,int}} + T_{\mathrm{I,last}} + T_{\mathrm{I,fin}} + T_{\mathrm{glob}} \\
&\quad + T_{\mathrm{pk},0} + (|\mathcal{P}|-k-|\mathcal{P}_{\mathrm{last}}|)\,\overline{T}_{\mathrm{pk,int}} + T_{\mathrm{pk,last}}.
\end{aligned}
\label{eq:app_proj}
\end{equation}
Because compiled bit-level Toffoli counts are deterministic and independent of random \gate{MX} outcomes, five stratified batches (batch~0, quartiles, and final batch, verified to contain maximal active widths and $t=\mathrm{ctz}(p-1)=1$ across our sorted prime pools) recover $\Qp$ with zero error and match all-prime Toffoli counts within $-0.10\%$ to $+0.02\%$ across all six certified circuits (Table~\ref{tab:all_prime}).

\needspace{5\baselineskip}
\subsection{Clean-ancilla accounting, bit-level gate costs, and matched \texorpdfstring{$b-2$}{b-2} conversion}
\label{app:lookup_regimes}

All circuits are profiled without dirty ancillas: every wire release verifies either deterministic reset to $\ket{0}$ or an \gate{MX} measurement paired with explicit phase cleanup. Peak logical qubits count live registers plus temporary clean ancillas at every execution step $\tau$:
\begin{equation}
\Qp = \max_{\tau}\bigl[Q_{\mathrm{live}}(\tau) + Q_{\mathrm{anc}}(\tau)\bigr].
\label{eq:qpeak}
\end{equation}

All \CmbNumConfigs{} circuits in our unified suite lower explicitly to elementary bit-level gates in the wire-stream format of Gu \emph{et al.}~\cite{gu2026}. Each forward clean-target \gate{AND} (costing four $T$ gates or one CCZ magic state with zero-$T$ measurement-based uncomputation~\cite{gidney2018halving}), \gate{CCX}, and \gate{CSWAP} counts as one strict bit-level Toffoli. Across all six certified circuits in Table~\ref{tab:hw_ext}(a), emitted non-Clifford streams comprise $94.0\%$--$99.0\%$ \gate{AND}, $1.0\%$--$2.0\%$ \gate{CCX}, and $0.1\%$--$4.0\%$ \gate{CSWAP}. Under CCZ-state teleportation as in Ref.~\cite{gidney2025}, \gate{CCX} and \gate{CSWAP} require zero extra data qubits, whereas pure four-$T$ decomposition~\cite{jones2013} without direct CCZ states adds $+1$ temporary ancilla wherever \gate{CCX} or \gate{CSWAP} executes at $\Qp$ on $w\in\{8,9\}$.

\leadin{Strict bit-level rules versus Gidney's macro model.}
Strict emitted bit-level counting charges the non-Clifford least-significant-bit (LSB) carry $x_0\wedge y_0$ on quantum-quantum adders and explicitly tracks even-constant $\mathrm{ctz}(c)$ savings, whereas Gidney's macro baseline~\cite{gidney2025} assigns simplified closed-form primitive costs:
\begin{center}
\footnotesize
\renewcommand{\arraystretch}{1.12}
\begin{tabular}{@{}lcc@{}}
\toprule
Primitive & Strict emitted bit-level (this work) & Gidney (2025) macro model~\cite{gidney2025} \\
\midrule
$b$-bit clean-carry adder & $b-1$ \gate{AND} gates (quantum--quantum); $\max(0,b-2-\mathrm{ctz}(c))$ (const.\ $c\neq 0$) & $\max(0,b-2)$ \\
$b$-bit comparator & Eliminated in dyadic mod-$2^f$ CRT & $b$ \\
$a$-bit QROM / ANF tree & $\max(0,2^a-a-1)$ \gate{AND} gates & $\max(0,2^a-a-2)$ \\
$K$-entry phase lookup & Explicit Half-DFS / ANF Clifford stream & $\lfloor\sqrt{K}\rfloor$ \\
\bottomrule
\end{tabular}
\end{center}
To place our bit-level frontier on equal adder footing with Ref.~\cite{gidney2025}, we compute the matched-adder reference
\begin{equation}
\Ts^{\mathrm{SA}} = \Ts - |\mathcal{P}|\bigl(W + c_{\mathrm{pk}}\bigr),
\label{eq:app_sa}
\end{equation}
where $c_{\mathrm{pk}} = 0$ for low-workspace $w=6$ schedules (whose Stages~(iii)--(iv) use 2-ancilla clean-\gate{AND} Cuccaro adders costing $2(b-1)$ \gate{AND} gates per addition, so no $b-2$ adjustment is applied) and $c_{\mathrm{pk}} = 35$ for standard schedules ($27$ in Stage~(iii) forward plus reverse and $8$ in Stage-(iv) dyadic CRT).

\begin{remark}
Equation~\eqref{eq:app_sa} is a nominal post-compilation conversion from our emitted strict bit-level streams, matching compiler instrumentation up to the negligible ($<0.1\%$) $+kW$ startup/uncompute boundary and $g_p=3$ window-0 terms; all \CmbNumPareto{} Pareto-optimal circuits use $c_{\mathrm{pk}}\in\{0,35\}$.
\end{remark}

\needspace{5\baselineskip}
\subsection{Invariance and transfer to arbitrary Eker{\aa}--H{\aa}stad two-base instances}
\label{app:twobase}

Following Ref.~\cite{gidney2025}, our concrete RSA-2048 gate-stream benchmarks fix $N$ to the RSA-2048 challenge integer ($N=251959084756\dots 0720357$) and set $a_{\mathrm{base}}=3$, $b_{\mathrm{base}}=a_{\mathrm{base}}^{2^{m_1}}\bmod N$, reducing the evaluation to the single-base proxy $3^x\bmod N$. Replacing this proxy by an arbitrary Eker{\aa}--H{\aa}stad two-base instance $g^{e_1}h^{-e_2}\bmod N$ (i.e., $a_{\mathrm{base}}=g$ and $b_{\mathrm{base}}=h^{-1}\bmod N$ over independent registers $e_1,e_2$ of lengths $m_1,m_2$) cleanly separates strictly invariant circuit properties from instance-dependent compile-time parameters.

Classical precomputation populates each $2^{w_j}$-entry window table $F_j(x_j)\in[1,N-1]$ as
\[
F_j(x_j) =
\begin{cases}
g^{2^{jw}x_j}\bmod N, & \text{on pure $e_1$-windows } (j < \lfloor m_1/w\rfloor),\\[2pt]
g^{2^{j^* w}x_{j^*,\mathrm{lo}}}h^{-x_{j^*,\mathrm{hi}}}\bmod N, & \text{on the straddling window } j^* = \lfloor m_1/w\rfloor \text{ when } w\nmid m_1,\\[2pt]
h^{-2^{jw-m_1}x_j}\bmod N, & \text{on pure $e_2$-windows } (j \ge \lceil m_1/w\rceil),
\end{cases}
\]
where, whenever $w\nmid m_1$, the straddling $w$-bit window splits as $x_{j^*}=x_{j^*,\mathrm{lo}}+2^{r^*}x_{j^*,\mathrm{hi}}$ with $r^*=m_1-j^*w\in\{1,\dots,w-1\}$ ($x_{j^*,\mathrm{lo}}\in\{0,\dots,2^{r^*}-1\}$ from $e_1$ and $x_{j^*,\mathrm{hi}}\in\{0,\dots,2^{w-r^*}-1\}$ from $e_2$). Even this single cross-boundary window retains identical width $w$, so the quantum circuit never inspects whether a window originates from $e_1$, $e_2$, or their boundary.

\leadin{Strictly invariant circuit properties.}
Because the circuit consumes only the resulting per-prime discrete-logarithm tables, the register topology (Fig.~\ref{fig:circuit_overview}), wire-lifecycle schedule, peak logical qubit count $\Qp$, Stage-(ii)--(iv) per-prime kernels, Stage-(i) QROM/ANF topologies, structural proofs (Propositions~1--5), and deterministic \gate{MX} invariants (Appendix~\ref{app:kickmix_theory}) remain strictly invariant. The single-run gate count $\Ts$ is likewise unchanged apart from two negligible effects: (i)~removing the $<0.01\%$ window-0 collapse on primes whose smallest primitive root happens to be $g_p=3$ under $a_{\mathrm{base}}=3$ (which shifts full-prime retention by $<3\times 10^{-6}$), and (ii)~negligible ($<0.1\%$) statistical fluctuations in prime count $|\mathcal{P}|$ and $\mathrm{ctz}(p-1)$ distribution after prime filtering.

\leadin{Instance-dependent compile-time steps.}
Only two classical steps recompute per instance $(N,g,h)$ in seconds:
\begin{enumerate}
\item \emph{Prime filtering.} Discarding primes dividing any table entry $F_j(x_j)$ leaves a survival fraction of $\approx 1-W(2^w-1)/p$ and a nearly identical prime count $|\mathcal{P}|$ for generic bases, while enforcing $L\ge N^W$ and $|L\bmod^{\pm}N|<N/2^{g_{\mathrm{gap}}}$.
\item \emph{Overflow tail certification.} For $D=27\mathrm{u}$ lanes (including our $w=9$ sub-billion endpoints and $s=8,w=8$ schedules), the exact rational Hoeffding bound (Appendix~\ref{app:mgf}, Eq.~\eqref{eq:app_hoeff}) depends only on $d_{p,j}\in[0,p-2]$ and $W$, and therefore holds universally for every 2048-bit modulus $N$ and every base pair $(g,h)$ without inspecting table distributions. For tighter $D=26\mathrm{u}$ lanes, the float-free integer MGF certificate depends on the realized discrete-logarithm distributions $\{d_{p,j}(x_j)\}$ and recomputes in seconds per instance at compile time; any prime exhibiting an anomalous distribution under a given $(g,h)$ is trivially discarded during prime filtering at negligible cost.
\end{enumerate}

\needspace{5\baselineskip}
\subsection{Dyadic error budget and 28-circuit Gidney baseline sweep}
\label{app:baseline_sweep}

For error-budget equivalence, our baseline sweep and \CmbNumParetoEqBudget{} of our \CmbNumPareto{} compiled Pareto circuits fix Gidney's parameters $(f=33,\mb=26,g_{\mathrm{gap}}=33)$, where initializing the low $\mb$ accumulator bits in a uniform coset-mask superposition~\cite{gidney2019,gidney2025} partitions $\mathbb{T}=\mathbb{R}/\mathbb{Z}$ into $2^{f-\mb}$ bins and $g_{\mathrm{gap}}$ guard bits enforce $|L\bmod^{\pm}N| < N/2^{g_{\mathrm{gap}}}$. Two wide-window $w=9$ endpoints at $2{,}823$ and $2{,}919$ qubits evaluate $(f=34,\mb=27,g_{\mathrm{gap}}=34)$, preserving $f-\mb=7$ at one additional qubit alongside an equal-budget $f=33$ control at $2{,}822$ qubits.

Retaining good shots across retries gives expected factoring Toffolis $\Tf = (s+1)\Ts / p_{\mathrm{keep}}$~\cite{ekera2020,gidney2025}. Under our rounded dyadic CRT modulo $2^f$ on the torus (Sec.~\ref{sec:align_repr} and Proposition~3, Appendix~\ref{app:prop3_dyadic}), integer wraparound multiples cancel modulo $1$ and symmetric rounding restricts each slice to diameter $2^{-f}$, yielding
\begin{equation}
\Tf = \frac{(s+1)\,\Ts}{p_{\mathrm{keep}}}, \qquad
p_{\mathrm{keep}}^{\mathrm{dyadic}} = 1 - 2^{\mb-f} - (s+1)\eta - 2^{f-\mb} M_{\mathrm{CRT}}\bigl(2^{-f}+2^{-g_{\mathrm{gap}}}\bigr),
\label{eq:tfact_dyadic}
\end{equation}
under the nonwrapping condition $e_{\max}-e_{\min} < 2^{\mb-f}$ (worst-case torus error diameter; Proposition~3), where $\eta \ge 2\sqrt{p_{\mathrm{bad}}}$ bounds the single-run trace distance from compressed Stage-(i) wraparound (Appendix~\ref{app:mgf}; $\eta=0$ on uncompressed lanes). We distinguish rigorously proven circuit error bounds from the factoring success estimator: Propositions~1--5 and Appendix~\ref{app:mgf} unconditionally certify the deterministic CRT rounding diameter $e_{\max}-e_{\min}$ and single-run overflow trace distance $\eta$, whereas translating those bounds into expected factoring Toffolis $\Tf$ via Eq.~\eqref{eq:tfact_dyadic} adopts Gidney's coset-mask retention and Eker{\aa}--H{\aa}stad postprocessing model (Assumption~2 of Ref.~\cite{gidney2025}) symmetrically across both candidate and baseline frontiers.

\leadin{Comparison across CRT formulations.}
In Gidney's baseline modulo $N_{\mathrm{trunc}}$~\cite{gidney2025}, floor truncation and wraparound mismatch double the per-slice truncation term from $2^{-f}$ to $2\cdot 2^{-f}$, so dyadic CRT cuts the worst-case per-slice error diameter by one-third at equal $M_{\mathrm{CRT}}$ and $g_{\mathrm{gap}}=f$. Although fusing Stage~(iv) into 3-bit slices doubles the slice count relative to unfused 5-bit CRT modulo $N_{\mathrm{trunc}}$, dyadic rounding recovers two-thirds of the resulting diameter increase on our primary $(s=4,|\mathcal{P}|=19{,}878,f=33,\mb=26)$ family:
\begin{center}
\footnotesize
\renewcommand{\arraystretch}{1.12}
\begin{tabular}{@{}lcccc@{}}
\toprule
CRT formulation & Slice count $M_{\mathrm{CRT}}$ & Worst-case diameter & Retention $p_{\mathrm{keep}}$ & Borrow / comparator \\
\midrule
Unfused 5-bit $\bmod\,N_{\mathrm{trunc}}$~\cite{gidney2025} & $4|\mathcal{P}| = 79{,}512$ & $2.78\times 10^{-5}$ & $0.9886$ ($\eta=0$) & Required \\
Fused 3-bit $\bmod\,N_{\mathrm{trunc}}$ & $8|\mathcal{P}| = 159{,}024$ & $5.55\times 10^{-5}$ & $0.9849$ (incl.\ $5\eta$) & Required \\
Fused 3-bit dyadic $\bmod\,2^f$ (this work) & $8|\mathcal{P}| = 159{,}024$ & $3.70\times 10^{-5}$ & $0.9873$ (incl.\ $5\eta$; $0.9874$ at $\eta=0$) & Eliminated \\
\bottomrule
\end{tabular}
\end{center}
Thus dyadic CRT trades only $0.13\%$ in full-run retention probability relative to unfused 5-bit CRT while cutting Stage-(iv) Toffolis by $45\%$ and removing the borrow qubit. Every $\Tf$ value reported in this work uses the exact $M_{\mathrm{CRT}}=8|\mathcal{P}|$ retention factor.

\leadin{Effect of the $0.99$ lattice-postprocessing factor.}
Gidney's constant $0.99$ classical lattice-postprocessing factor in the denominator of $\Tf$ is omitted symmetrically on both candidate and baseline frontiers. Restoring $0.99$ keeps every sub-billion milestone below $10^9$ under both matched accounting ($0.9882\times 10^9$ at $2{,}433$ qubits, $0.9705\times 10^9$ at $2{,}485$ qubits, and $0.9411\times 10^9$ at $2{,}919$ qubits) and strict emitted counting ($0.9936\times 10^9$ at $2{,}485$ qubits, $0.9898\times 10^9$ at $2{,}822$ qubits, $0.9881\times 10^9$ at $2{,}823$ qubits, and $0.9592\times 10^9$ at $2{,}919$ qubits).

\leadin{28-circuit Gidney (2025) baseline sweep.}
To compare against a full baseline frontier under uniform clean-ancilla accounting, we evaluated Gidney's unchanged 2025 serial RNS architecture~\cite{gidney2025} across \NumRefValid{} valid parameter tuples at $(f=33,\mb=26,g_{\mathrm{gap}}=33)$ spanning $s\in\{2,\dots,14\}$, $\ell\in\{19,20,21\}$, and $w\in\{6,7,8\}$. Its \NumRefFrontier-point Pareto frontier runs from $1{,}323$ qubits at $(14,19,7)$ ($10.96\times 10^9$ Toffolis) to $1{,}863$ qubits at $(3,21,6)$ ($4.487\times 10^9$ Toffolis), reaching its minimum qubit--Toffoli product $\Qp\Tf=\RefMinProdVal$ at $(4,21,6)$ ($1{,}691$ qubits, $4.705\times 10^9$ Toffolis). Under explicit clean-ancilla tracking, Gidney's published $(8,21,6)$ configuration uses $1{,}435$ qubits and $6.388\times 10^9$ Toffolis, versus $1{,}399$ qubits from the closed-form formula omitting $36$ temporary ancillas.

\needspace{12\baselineskip}
\section{LLM-Driven Compilation and Dual-Backend Verification}
\label{app:verif}

This appendix documents how our circuits were discovered and independently certified at full RSA-2048 width. Appendix~\ref{app:kickmix_theory} formalizes why Shor's modular-exponentiation core admits deterministic classical trajectory tracking and defines our symbolic and randomized \emph{kickmix} backends; Appendix~\ref{app:workflow} describes the researcher-frozen LLM search loop and two-tier screening protocol; and Appendix~\ref{app:calibration} summarizes the 330-bit multi-segment validations and full-width certification across all primes (Table~\ref{tab:all_prime}).

\needspace{5\baselineskip}
\subsection{Deterministic \gate{MX} invariants and symbolic all-\texorpdfstring{$2^M$}{2\^M} tracking}
\label{app:kickmix_theory}

Full state-vector simulation is intractable beyond $\sim 40$ qubits, yet Shor's modular-exponentiation core admits exact classical trajectory verification. Because outer Hadamards and the terminal inverse QFT lie outside the arithmetic core, every internal gate is a basis permutation, a diagonal $\mathbb{F}_2$ phase gate, or an \gate{MX} measurement on a wire holding a deterministic Boolean function of surviving computational-basis registers, with classical \gate{MX} outcomes controlling only diagonal Clifford phase gates and never basis permutations. Consequently, distinct computational-basis branches never merge, and the accumulated phase remains affine-linear in the $M$ mid-circuit outcomes $\{m_\nu\}$.

\needspace{6\baselineskip}
Inside our dyadic circuits, every \gate{MX}-measured wire belongs to one of three deterministic classes:
\begin{enumerate}
\item Temporary \gate{AND} ancillas whose two control wires remain live and unmodified during \gate{MX}$+$\gate{CZ}.
\item QROM data outputs $T(a)$ measured while the address register $a$ remains live during phase cleanup.
\item Early-measured Stage-(iii) mod-$p$ words determined uniquely by the live reduced exponent $r_p(x)$ (the truncated-modulus borrow wire is absent in dyadic mod-$2^f$ CRT).
\end{enumerate}

\leadin{Dual-backend verification engines.}
To certify the $3.14\text{ TB}$ of full-width gate streams, every binary segment was cross-checked with two independent compiled backends, namely the upstream \emph{andigu/uncompute} C backend of Gu \emph{et al.}~\cite{gu2026} executing randomized \emph{kickmix} replay~\cite{khattar2026}, and our symbolic parser over all $2^M$ measurement outcomes:
\begin{itemize}
\item \emph{Symbolic all-$2^M$-outcome simulator.} Each \gate{MX} measurement $\nu\in\{0,\dots,M-1\}$ allocates an independent $\mathbb{F}_2$ indeterminate $m_\nu$, and each derived classical parity handle $c$ stores its sparse support $\mathrm{supp}(c)\subseteq\{0,1,\dots,M\}$ ($0$ encoding constant phase $-1$ and $\nu+1$ encoding $m_\nu$). On each trajectory $\tau$, the simulator maintains the accumulated phase as a sparse set $P_\tau\subseteq\{0,1,\dots,M\}$: when \gate{MX} $\nu$ measures a wire in state $\ket{1}$, it toggles $P_\tau \leftarrow P_\tau \triangle \{\nu+1\}$ (unchanged for $\ket{0}$), and each active global sign-flip, \gate{Z}, or \gate{CZ} gate updates $P_\tau \leftarrow P_\tau \triangle \mathrm{supp}(c)$ over $\mathbb{F}_2$. Checking $P_\tau=\emptyset$ at exit proves zero residual phase simultaneously across all $2^M$ measurement bitstrings with zero retained memory per retired outcome.
\item \emph{Randomized kickmix replay.} Strengthening Ref.~\cite{khattar2026}, an independent pass evaluates the same stream with SHAKE-256 measurement bits seeded by the stream SHA-256 digest and domain nonce, cross-checking basis outputs, $\Qp$, total Toffolis, and zero residual phase.
\end{itemize}

\begin{remark}[Verification boundaries]
Two scope boundaries remain explicit: coverage across $2^{m_e+\mb}$ computational-basis inputs is sampled over $32$ trajectories per certified circuit, while rare wrapped-register tails ($p_{\mathrm{bad}}\le 2.83\times 10^{-10}$ at $s=4,w=8,D=26\mathrm{u}$) are certified analytically by Propositions~1--2 and Appendix~\ref{app:mgf}.
\end{remark}

\needspace{5\baselineskip}
\subsection{LLM-driven compilation workflow and two-tier screening}
\label{app:workflow}

Because all-prime RSA-2048 replay takes tens of minutes even on compiled C++ backends, our compilation workflow separates fast Tier-1 search screening from exhaustive Tier-2 gate-stream certification. Researchers freeze the objective, approximation margins, cost conventions, and independent verification checks outside the LLM search loop's modification scope, reviewing peak-qubit and per-prime diagnostics periodically between rounds. During our 2026 discovery rounds, GPT-6 Astra operated inside this search loop to generate offset unsigned lanes, non-restoring conjugated division, dyadic torus CRT, rolling ANF, and Half-DFS phase schedules.

\leadin{Tier-1 admission gates.}
Every candidate edit was admitted only after passing three automated checks:
\begin{enumerate}
\item \emph{Small-width exhaustive and fault-injection tests}, including wrapped Stage-(ii) states $U\ge 2H$ across $D\in\{8,9,10,11\}$ and $t\in\{1,\dots,4\}$.
\item \emph{Scaled 330-bit simulations across all primes}, testing nondivisible partial batches ($|\mathcal{P}|\bmod k\neq 0$), clean chaining across multiple segments, and stratified batch calibration within $1\%$.
\item \emph{Full-width RSA-2048 profiling} over five stratified batches paired with analytical tail certificates (Appendix~\ref{app:layout_proj}).
\end{enumerate}

\leadin{Illustrative Tier-1 catches.}
Two concrete catches during search demonstrate why independent symbolic and nondivisible checks were essential:
\begin{itemize}
\item \emph{Omitted constant-term phase flip.} An early rolling-ANF uncomputation candidate produced exact computational-basis outputs and clean ancilla resets yet omitted the constant-term ($A=\emptyset$) global sign flip. Tier-1 symbolic tracking immediately flagged $P_\tau\neq\emptyset$, prompting the $\langle u_{\mathrm{out}},a_\emptyset\rangle$ sign-flip fix (Proposition~4(b)) that restored $P_\tau=\emptyset$ across all $2^M$ outcomes.
\item \emph{Masked empty-slot boundary bug.} An empty-slot bug in partial final batches was masked on a 330-bit $s=4,w=8$ test whose $624=48\times 13$ primes divided evenly by $k=13$, and was caught immediately once nondivisible ($|\mathcal{P}|\bmod k\neq 0$) checks were enforced.
\end{itemize}

At convergence, Tier~2 partitions full-width circuits into clean batch segments for dual-backend replay across all primes. All final counts and verifications reproduce deterministically from the released compiler and inputs without re-running the LLM search loop.

\needspace{5\baselineskip}
\subsection{330-bit multi-segment chains and full-width certification across all primes}
\label{app:calibration}

\leadin{Scaled 330-bit multi-segment chains.}
On scaled 330-bit moduli ($527$--$683$ primes at $\ell_{\mathrm{val}}=17$), three multi-batch chains split across two segments passed both symbolic all-$2^M$ and randomized \emph{kickmix} verification across four input pairs: a $660$-qubit rolling ANF chain (zero deferred phase), a $484$-qubit zero-tail packed buffer chain (explicit cross-batch $\mathbb{F}_2$ XOR phase accumulation), and a $325$-qubit low-workspace bilinear ANF chain (zero deferred phase). An additional 624-prime numerical dyadic CRT oracle confirmed a maximum torus error of $6.66\times 10^{-8}$.

\leadin{Full-width RSA-2048 certification and same-RNS control.}
At full RSA-2048 width, Table~\ref{tab:all_prime} summarizes dual-backend certification across all constituent primes and $32$ trajectories ($16$ primary plus $16$ seed-isolated holdout) for six representative frontier circuits, totaling $119{,}482$ constituent 20-bit prime evaluations across the six circuits, $6{,}505$ clean batch segments ($5{,}610$ with zero segmentation overhead), and $3.14\text{ TB}$ of unique bit-level streams. Three additional milestone circuits at $1{,}359$, $2{,}822$, and $2{,}919$ qubits passed dual-backend verification on batch~0, confirming $\Qp$, basis outputs, clean ancilla reset, and zero symbolic phase across all $2^M$ outcomes, alongside exact monolithic counts across all primes at $2{,}822$ and $2{,}919$ qubits. Finally, for the same-RNS controlled progression in Fig.~\ref{fig:ablations}(b) and Sec.~\ref{sec:case_study}, the bit-level mod-$N_{\mathrm{trunc}}$ baseline at $(s=4,k=13,w=8,f=33)$ ($2{,}204$ qubits, $246.97\times 10^6$ Toffolis per run, $3.23\times 10^6$ segmentation overhead) passed dual-backend verification across all $19{,}878$ primes over $16$ trajectories.

\begin{table}[!htbp]
\caption{All-prime, $32$-trajectory bit-level gate-stream certification for six representative frontier circuits. ``Seg.\ $\Delta T = T_{\mathrm{seg}}-T_{\mathrm{mono}}$'' counts tagged boundary phase flushes ($0$ across all rolling and bilinear segments); ``Max torus err.'' reports the largest $|z/2^f-(3^x\bmod N)/N|_{\mathbb{T}}$ over $32$ trajectories (vs.\ Prop.~3 bounds $2.9\times 10^{-5}$ at $f=33$ and $1.4\times 10^{-5}$ at $f=34$); ``Strat.\ $\Delta$'' gives the 5-batch projection deviation relative to exact all-prime measurement.}
\label{tab:all_prime}
\centering
\footnotesize
\renewcommand{\arraystretch}{1.08}
\begin{tabular*}{\textwidth}{@{\extracolsep{\fill}}lccrrrrrr@{}}
\toprule
Certified circuit & $(s,k,w,D,f)$ & Primes / segs & $\Qp$ & $T_{\mathrm{mono}}$ & Seg.\ $\Delta T$ & $\Tf$ ($10^9$) & Max torus err. & Strat.\ $\Delta$ \\
\midrule
Bilinear low-workspace ($k{=}1$) & $(14,1,6,26\mathrm{u},33)$ & $20{,}288 / 1{,}353$ & $1{,}306$ & $417{,}843{,}798$ & $0$ & $6.3548$ & $7.37\times 10^{-8}$ & $-0.10\%$ \\
Rolling ANF ($k{=}13$) & $(4,13,8,26\mathrm{u},33)$ & $19{,}878 / 1{,}530$ & $2{,}199$ & $227{,}414{,}986$ & $0$ & $1.1517$ & $7.62\times 10^{-8}$ & $+0.02\%$ \\
Rolling ANF ($k{=}20$) & $(4,20,8,26\mathrm{u},33)$ & $19{,}878 / 994$ & $2{,}381$ & $202{,}028{,}097$ & $0$ & $1.0232$ & $8.23\times 10^{-8}$ & $+0.01\%$ \\
Rolling ANF ($k{=}22$, min $\Qp\Tf$) & $(4,22,8,26\mathrm{u},33)$ & $19{,}878 / 904$ & $\mathbf{2{,}433}$ & $197{,}690{,}762$ & $0$ & $\mathbf{1.0012}$ & $7.37\times 10^{-8}$ & $-0.03\%$ \\
Rolling ANF ($k{=}24$, strict $<10^9$) & $(4,24,8,26\mathrm{u},33)$ & $19{,}878 / 829$ & $2{,}485$ & $194{,}222{,}346$ & $0$ & $0.9836$ & $8.06\times 10^{-8}$ & $-0.02\%$ \\
Wide-window packed ($k{=}22$) & $(3,22,9,27\mathrm{u},34)$ & $19{,}682 / 895$ & $2{,}823$ & $242{,}077{,}732$ & $6{,}261{,}576$ & $0.9782$ & $4.51\times 10^{-8}$ & $+0.01\%$ \\
\bottomrule
\end{tabular*}
\end{table}

\needspace{12\baselineskip}
\section{Mathematical Proofs and Tail-Bound Certificates}
\label{app:proofs}

This appendix establishes the formal correctness, resource complexity, and error bounds behind the compilation mechanisms of Sec.~\ref{sec:principles}. Appendices~\ref{app:prop1}--\ref{app:prop5_trunc} prove Propositions~1--5 covering offset unsigned encoding, non-restoring $\mathrm{ctz}(p-1)$-sliced division (summarized in Algorithm~1), dyadic torus CRT rounding, localized ANF/Half-DFS phase cleanup, and sign-extension-free lane addition; Appendix~\ref{app:mgf} derives the fixed-point MGF, rational Hoeffding, and Stinespring trace-distance bounds certifying compressed lanes beyond trajectory sampling.

\needspace{5\baselineskip}
\subsection{Offset unsigned encoding and wrapped-branch reset}
\label{app:prop1}

\begin{proposition}[Offset unsigned encoding and wrapped-branch reset]
For an $\ell$-bit odd prime $p\in(2^{\ell-1},2^\ell)$ and lane width $D>\ell$, set $H = 2^{D-\ell-1}(p-1)$, $\mu = \lfloor\mathbb{E}[S]\rceil$, and $B = \mu - H$, storing $U(x) = (S(x)-B)\bmod 2^D$. Then:
\begin{enumerate}
\item[(i)] for every nonoverflowing input $S(x)\in[\mu-H,\mu+H)$, in-place reduction on the $D$ wires computes $r=U(x)\bmod(p-1)$ with zero extra persistent qubits and Stage~(iii) evaluates $g^{r+B}\equiv g^{S(x)}\pmod p$; and
\item[(ii)] for every input $x$, including wrapped branches $S(x)\notin[\mu-H,\mu+H)$, forward and reverse Stage~(ii) followed by telescoping batch transitions restore all Stage-(i) lanes and Stage-(iii) scratch wires cleanly to $\ket{0}$.
\end{enumerate}
\end{proposition}

\begin{proof}
\proofstep{Nonoverflowing reduction.}
Whenever $S(x)\in[\mu-H,\mu+H)=[B,B+2H)$, the shifted integer $U(x)=S(x)-B\in[0,2H)$ satisfies $U(x)<2H=2^{D-\ell}(p-1)<2^D$. In-place binary reduction on the $D$ wires therefore computes $r = U\bmod(p-1)$ in wires $0\dots\ell-1$ and complemented quotient bits $\overline{q}$ in wires $\ell\dots D-1$ with zero extra persistent qubits. Precomputing the Stage-(iii) startup table as $g^{r_{0\dots b_0-1}+B}\bmod p$ ensures $g^{r+B}\equiv g^{U+B}\equiv g^S\pmod p$ by Fermat's little theorem.

\proofstep{Batch-to-batch offset updates.}
Each batch transition (including batch~0 initialization and final lane clear) applies $(B_{\mathrm{old}}-B_{\mathrm{new}})\bmod 2^D$ via one unconditional $D$-bit constant adder, taking $B_{\mathrm{old}}=0$ on batch~0 and $B_{\mathrm{new}}=0$ at exit. On batch~0, $-B\bmod 2^D$ can alternatively be folded into zero-Toffoli \gate{X} gates prior to window addition.

\proofstep{Wrapped-branch reversibility and clean reset.}
On rare wrapped branches $S(x)\notin[B,B+2H)$, forward Stage~(ii) $R_{\mathrm{div}}$ acts as a bijection on $\{0,1\}^D$ (Proposition~2). Stages~(iii)--(iv) read $r$ without mutating those $D$ wires, and every Stage-(iii) lookup entry is pre-reduced modulo $p$ for all $r\in\{0,1\}^\ell$, so intermediate sums stay within $[0,8(p-1)]\subset[0,2^{\ell+3})$ at $\ell=20$ and every Stage-(iii) scratch wire resets cleanly to $\ket{0}$. Hence $R_{\mathrm{div}}^\dagger$ restores $\ket{U_{\mathrm{old}}(x)}$ identically on every branch, and telescoping additions $(U_{\mathrm{new}}-U_{\mathrm{old}})\bmod 2^D$ in $(\mathbb{Z}/2^D\mathbb{Z},+)$ guarantee final lane uncomputation to $\ket{0}^{\otimes D}$ on wrapped and nonwrapped branches alike.
\end{proof}

\needspace{5\baselineskip}
\subsection{Non-restoring \texorpdfstring{$\mathrm{ctz}(p-1)$}{ctz(p-1)}-sliced conjugated division}
\label{app:prop2_loop2}

\begin{proposition}[Non-restoring $\mathrm{ctz}(p-1)$-sliced conjugated division]
Let $p\in(2^{\ell-1},2^\ell)$ be an odd prime and $D>\ell$ be the lane width. Write $p-1 = 2^t m_{\mathrm{odd}}$ with $t=\mathrm{ctz}(p-1)\ge 1$ and $m_{\mathrm{odd}}$ odd, and set $k_{\mathrm{div}}=D-\ell$ and $w_{\mathrm{div}}=\ell+1-t$. Then Algorithm~1 maps every $U\in[0,2^{k_{\mathrm{div}}}(p-1))$ in place to $(r,\overline{q})$ with $r=U\bmod(p-1)$ and acts as a bijection on all $\{0,1\}^D$, using $2(k_{\mathrm{div}}+1)\max(0,w_{\mathrm{div}}-2)$ clean-target \gate{AND} gates forward plus reverse and zero additional persistent qubits.
\end{proposition}

\begin{center}
\begin{minipage}{0.94\textwidth}
\footnotesize
\noindent\rule{\linewidth}{0.8pt}\\[2pt]
\textbf{Algorithm 1:} In-place non-restoring $\mathrm{ctz}(p-1)$-sliced conjugated division $R_{\mathrm{div}}$\\[-5pt]
\noindent\rule{\linewidth}{0.4pt}\\[3pt]
\textbf{Input:} $D$-bit register $U_{0\dots D-1}$; odd prime $p\in(2^{\ell-1},2^\ell)$ with $p-1=2^t m_{\mathrm{odd}}$, $k_{\mathrm{div}}=D-\ell$, $w_{\mathrm{div}}=\ell+1-t$.\\[1.5pt]
\textbf{Output:} In-place $(r,\overline{q})$ with $r=U\bmod(p-1)$ on wires $0\dots\ell-1$ and $\overline{q}$ on wires $\ell\dots D-1$ whenever $U<2^{k_{\mathrm{div}}}(p-1)$.
\begin{enumerate}[label=\textbf{\arabic*.},leftmargin=1.8em,itemsep=1.5pt,topsep=3pt]
\item \textbf{Initial subtraction} ($j=k_{\mathrm{div}}-1$): subtract odd constant $m_{\mathrm{odd}}$ modulo $2^{w_{\mathrm{div}}}$ on slice $U_{j+t\dots D-1}$; top wire holds $\mathrm{sign}_{k_{\mathrm{div}}-1}=\overline{q}_{k_{\mathrm{div}}-1}$.
\item \textbf{for} $j = k_{\mathrm{div}}-2$ \textbf{down to} $0$ \textbf{do}
\item \hangindent=1.2em\hspace{1.2em}Let $X \leftarrow U_{j+t\dots j+t+w_{\mathrm{div}}-1}$ and $\sigma \leftarrow \mathrm{sign}_{j+1}$ (wire $j+t+w_{\mathrm{div}}$).
\item \hangindent=1.2em\hspace{1.2em}Apply \gate{X}$(\sigma)$, transversal \gate{CX}$(\sigma, X_i)$ for $i=0\dots w_{\mathrm{div}}-1$, \emph{uncontrolled} $+m_{\mathrm{odd}}\pmod{2^{w_{\mathrm{div}}}}$ on $X$, transversal \gate{CX}$(\sigma, X_i)$, and \gate{X}$(\sigma)$.
\item \textbf{end for}
\item \textbf{Final restoration}: apply $\mathrm{sign}_0$-controlled $+m_{\mathrm{odd}}\pmod{2^{w_{\mathrm{div}}-1}}$ on slice $U_{t\dots \ell-1}$.
\end{enumerate}
\vspace{-4pt}\noindent\rule{\linewidth}{0.8pt}
\end{minipage}
\end{center}

\begin{proof}
\proofstep{Window narrowing and complement identity.}
Since $p<2^\ell$ is odd, $t\ge 1$, $w_{\mathrm{div}}\le \ell$, and $m_{\mathrm{odd}}=(p-1)/2^t < 2^{w_{\mathrm{div}}-1}$. Because $(p-1)2^j = m_{\mathrm{odd}} 2^{j+t}$ has $j+t$ trailing zeros, adding or subtracting $(p-1)2^j$ leaves the lowest $j+t$ wires invariant, narrowing each step window to $w_{\mathrm{div}}$ bits starting at bit $j+t$. For any $w_{\mathrm{div}}$-bit unsigned integer $X$, bitwise complement $\overline{X}=(2^{w_{\mathrm{div}}}-1)-X$ satisfies
\begin{equation}
\overline{\vphantom{\big(}(\overline{X}+m_{\mathrm{odd}})\bmod 2^{w_{\mathrm{div}}}} \equiv X - m_{\mathrm{odd}} \pmod{2^{w_{\mathrm{div}}}}.
\end{equation}

\proofstep{Inductive remainder invariant.}
For any valid input $U\in[0,2^{k_{\mathrm{div}}}(p-1))$, let $R^{(j)} = \lfloor U^{(j)} / 2^{j+t} \rfloor$ denote the shifted signed partial remainder on the active $w_{\mathrm{div}}$-bit window after step $j$. At $j=k_{\mathrm{div}}-1$, subtracting $m_{\mathrm{odd}}$ from $\lfloor U / 2^{k_{\mathrm{div}}-1+t} \rfloor \in [0, 2m_{\mathrm{odd}}-1]$ gives $R^{(k_{\mathrm{div}}-1)} \in [-m_{\mathrm{odd}}, m_{\mathrm{odd}}-1]$. Inductively, doubling and appending bit $j+t$ gives
\[
2R^{(j+1)} + u_{j+t} \in [-2m_{\mathrm{odd}},\, 2m_{\mathrm{odd}}-1];
\]
subtracting $m_{\mathrm{odd}}$ when $R^{(j+1)}\ge 0$ or adding $m_{\mathrm{odd}}$ when $R^{(j+1)}<0$ preserves $R^{(j)}\in[-m_{\mathrm{odd}},\,m_{\mathrm{odd}}-1]$. Since $m_{\mathrm{odd}} < 2^{w_{\mathrm{div}}-1}$, $R^{(j)}$ fits in $w_{\mathrm{div}}$-bit two's complement without overflow, so its most-significant-bit sign wire $\mathrm{sign}_j$ encodes $q_j = \overline{\mathrm{sign}_j}$ unambiguously.

\proofstep{Conjugated implementation and \gate{AND} count.}
At each step $j=k_{\mathrm{div}}-2,\dots,0$, conjugating one \emph{uncontrolled} odd-constant adder $+m_{\mathrm{odd}}$ by \gate{X} on $\mathrm{sign}_{j+1}$ followed by transversal \gate{CX}$(\mathrm{sign}_{j+1}, X_i)$ evaluates Eq.~\eqref{eq:nonrestoring} while leaving $\mathrm{sign}_{j+1}$ unchanged. Because bit $0$ of $m_{\mathrm{odd}}$ is $1$ (no LSB carry \gate{AND}) and the top carry is discarded, each step costs $\max(0,w_{\mathrm{div}}-2)$ \gate{AND} gates. After $j=0$, a single $\mathrm{sign}_0$-controlled restoration $+m_{\mathrm{odd}}$ on the remaining $(w_{\mathrm{div}}-1)$-bit slice maps $[-m_{\mathrm{odd}},-1]$ into $[0,m_{\mathrm{odd}}-1]$ in $\max(0,w_{\mathrm{div}}-2)$ \gate{AND} gates (every carry satisfies $c_i \le \mathrm{sign}_0$, costing one \gate{AND} at bit $0$ and one for each bit $1,\dots,w_{\mathrm{div}}-3$ when $w_{\mathrm{div}}\ge 3$, and $0$ when $w_{\mathrm{div}}=2$). Summing initial subtraction, $k_{\mathrm{div}}-1$ conjugated steps, and final restoration gives $2(k_{\mathrm{div}}+1)\max(0,w_{\mathrm{div}}-2)$ \gate{AND} gates forward plus reverse.

\proofstep{Global bijectivity on $\{0,1\}^D$.}
Because every step composes \gate{X}, \gate{CX}, and modular adders on $\{0,1\}^D$, the forward map $R_{\mathrm{div}}$ is a bijection on all $2^D$ computational states including wrapped $U\ge 2^{k_{\mathrm{div}}}(p-1)$, so its inverse $R_{\mathrm{div}}^\dagger$ restores $\ket{U}$ unconditionally.
\end{proof}

\begin{remark}[Typical RSA-2048 gate cost]
For 20-bit odd primes ($\ell=20$), the trailing-zero count $t=\mathrm{ctz}(p-1)$ averages $\approx 2$, shrinking the active window width $w_{\mathrm{div}}=\ell+1-t$ from $21$ to an average of $19$ bits ($\approx 17$ \gate{AND} gates per step). At $D=26$ ($k_{\mathrm{div}}=6$), $2(k_{\mathrm{div}}+1)\max(0,w_{\mathrm{div}}-2)$ therefore averages $238$ Toffolis per prime across our $19{,}878$-prime pool, cutting standard restoring division ($504$ Toffolis) by more than half.
\end{remark}

\needspace{5\baselineskip}
\subsection{Dyadic torus \texorpdfstring{$\bmod\,2^f$}{mod 2\^f} rounded CRT}
\label{app:prop3_dyadic}

\begin{proposition}[Dyadic torus $\bmod\,2^f$ rounded CRT]
Let $L = \prod_{p\in\mathcal{P}} p = K_0 N + \delta$ with odd RSA modulus $N$, $|\delta| < N/2^{g_{\mathrm{gap}}}$, and $L \ge N^W$. For input exponent $x$, let $X_{\mathrm{prod}} = \prod_{j=0}^{W-1} F_j(x_j) \in [0,L)$ satisfy $X_{\mathrm{prod}} \equiv Y \pmod N$ with $Y \in [0,N)$. For each $\ell$-bit prime $p\in\mathcal{P}$ ($\ell=20$), let Stage~(iii) output a noncanonical intermediate sum $\widetilde{A}_p = \sum_{i=0}^{7} w_{p,i} 2^{3i} \in [0,8(p-1)] \subset [0,2^{23})$ with 3-bit slices $w_{p,i}\in\{0,\dots,7\}$, and let $g^{r_{\mathrm{tail}}2^{o_{\mathrm{tail}}}}\bmod p$ denote the terminal Stage-(iii) exponent factor fused into Stage~(iv), satisfying $\widetilde{A}_p g^{r_{\mathrm{tail}}2^{o_{\mathrm{tail}}}} \equiv X_{\mathrm{prod}} \pmod p$ (the canonical-residue case corresponds to $r_{\mathrm{tail}}=0$). Across the resulting $M_{\mathrm{CRT}}=8|\mathcal{P}|$ fused slices $\alpha=(p,i)$, define
\[
t_{p,i} = \bigl(w_{p,i} 2^{3i} g^{r_{\mathrm{tail}}2^{o_{\mathrm{tail}}}}\bigr)\bmod p,
\qquad
C'_{p,i} = \bigl((t_{p,i} u_p)\bmod p\bigr)\cdot(L/p)\in[0,L),
\qquad
C_\alpha = C'_\alpha \bmod N,
\]
with $u_p=(L/p)^{-1}\bmod p$, and define $v_\alpha$ by Eq.~\eqref{eq:dyadic_vm}. Then accumulating $Z_{\mathrm{out}} = (Z_{\mathrm{mask}} + \sum_{\alpha=1}^{M_{\mathrm{CRT}}} v_\alpha)\bmod 2^f$ requires zero borrow qubits and satisfies single-point torus error $|e(x)|_{\mathbb{T}} < M_{\mathrm{CRT}}(\tfrac{1}{2}2^{-f}+2^{-g_{\mathrm{gap}}})$ and worst-case diameter $e_{\max}-e_{\min} < M_{\mathrm{CRT}}(2^{-f}+2^{-g_{\mathrm{gap}}})$.
\end{proposition}

\begin{proof}
\proofstep{Fused CRT lifting identity.}
For each slice $\alpha=(p,i)$, $(t_{p,i}u_p)\bmod p \in \{0,\dots,p-1\}$ implies $C'_{p,i}\in[0,L)$. Modulo $p$,
\[
\sum_{i=0}^{7} C'_{p,i}
\equiv \sum_{i=0}^{7} t_{p,i} u_p (L/p)
\equiv \Bigl(\sum_{i=0}^{7} w_{p,i} 2^{3i}\Bigr) g^{r_{\mathrm{tail}}2^{o_{\mathrm{tail}}}}
\equiv \widetilde{A}_p g^{r_{\mathrm{tail}}2^{o_{\mathrm{tail}}}}
\equiv X_{\mathrm{prod}} \pmod p,
\]
whereas for every other prime $p'\in\mathcal{P}\setminus\{p\}$, $L/p\equiv 0\pmod{p'}$ gives $C'_{p,i}\equiv 0\pmod{p'}$. By the Chinese Remainder Theorem, $\sum_{\alpha=1}^{M_{\mathrm{CRT}}} C'_\alpha \equiv X_{\mathrm{prod}} \pmod L$. Since $X_{\mathrm{prod}}\in[0,L)$ and $0 \le \sum_\alpha C'_\alpha < M_{\mathrm{CRT}} L$, there exists an integer $q(x)\in\{0,\dots,M_{\mathrm{CRT}}-1\}$ such that $\sum_{\alpha=1}^{M_{\mathrm{CRT}}} C'_\alpha = X_{\mathrm{prod}} + q(x)L$. Different noncanonical slice decompositions generally yield different $q(x)$, rounding residuals $\varepsilon_\alpha$, and accumulator words $Z_{\mathrm{out}}$, but all satisfy the same uniform bound derived below. Substituting $L = K_0 N + \delta$ and reducing modulo $N$ gives the exact identity
\begin{equation}
\sum_{\alpha=1}^{M_{\mathrm{CRT}}} C_\alpha = Y + m_{\mathrm{wrap}}(x) N + q(x)\delta
\end{equation}
for some integer $m_{\mathrm{wrap}}(x)\in\mathbb{Z}$.

\proofstep{Symmetric dyadic rounding on $\mathbb{T}$.}
By integer division, $2^f C_\alpha + \lfloor N/2\rfloor = q_\alpha N + r_\alpha$ with $0\le r_\alpha \le N-1$ and $v_\alpha \equiv q_\alpha \pmod{2^f}$, so $q_\alpha N - 2^f C_\alpha = \lfloor N/2\rfloor - r_\alpha \in [-\tfrac{N-1}{2},\,\tfrac{N-1}{2}]$ for odd $N$. Dividing by $2^f N$ bounds each slice's torus rounding error $\varepsilon_\alpha := v_\alpha/2^f - C_\alpha/N \pmod 1$ by $\varepsilon_\alpha \in [-\tfrac{1}{2}2^{-f}, +\tfrac{1}{2}2^{-f}]$. Summing over $\alpha$ modulo $1$, the integer wrap multiple $m_{\mathrm{wrap}}(x)N/N \equiv 0 \pmod 1$ cancels identically on $\mathbb{T}=\mathbb{R}/\mathbb{Z}$, leaving an unwrapped torus representative $\tilde{e}(x)$ satisfying
\begin{equation}
e(x) := \frac{(Z_{\mathrm{out}}-Z_{\mathrm{mask}})\bmod 2^f}{2^f} - \frac{Y}{N} \equiv \tilde{e}(x) := \sum_{\alpha=1}^{M_{\mathrm{CRT}}} \varepsilon_\alpha + \frac{q(x)\delta}{N} \pmod 1.
\end{equation}

\proofstep{Worst-case diameter and retention probability.}
Here $\sum_\alpha \varepsilon_\alpha \in [-\tfrac{1}{2}M_{\mathrm{CRT}}2^{-f}, +\tfrac{1}{2}M_{\mathrm{CRT}}2^{-f}]$ has diameter $M_{\mathrm{CRT}}2^{-f}$, while the fixed-sign RNS drift $q(x)\delta/N$ lies in an interval of diameter at most $M_{\mathrm{CRT}}2^{-g_{\mathrm{gap}}}$. This yields $|e(x)|_{\mathbb{T}} \le |\tilde{e}(x)| < M_{\mathrm{CRT}}(\tfrac{1}{2}2^{-f} + 2^{-g_{\mathrm{gap}}})$ and $e_{\max} - e_{\min} := \max_x \tilde{e}(x) - \min_x \tilde{e}(x) < M_{\mathrm{CRT}}(2^{-f} + 2^{-g_{\mathrm{gap}}})$. Under the nonwrapping condition $e_{\max}-e_{\min} < 2^{\mb-f}$ and Gidney's coset-mask overlap analysis~\cite{gidney2019,gidney2025}, combining pairwise overlap $1 - 2^{f-\mb}(e_{\max}-e_{\min})$ with zero-frequency probability $2^{\mb-f}$ and conservatively subtracting the multi-run compression trace-distance bound $(s+1)\eta$ (by subadditivity over $R=s+1$ independent runs from Appendix~\ref{app:mgf}) yields Eq.~\eqref{eq:tfact_dyadic}. At equal $M_{\mathrm{CRT}}$ and $g_{\mathrm{gap}}=f$, this cuts the worst-case total error diameter by one-third, from $3M_{\mathrm{CRT}}2^{-f}$ to $2M_{\mathrm{CRT}}2^{-f}$ ($3\cdot 2^{-f}$ to $2\cdot 2^{-f}$ per slice), while removing the borrow qubit and comparator.
\end{proof}

\needspace{11\baselineskip}
\subsection{Half-DFS phase trees and ANF Clifford cleanup}
\label{app:prop4_phase}

\begin{proposition}[Half-DFS phase trees and ANF Clifford cleanup]
The following localized phase-cleanup guarantees hold:
\begin{enumerate}
\item[(a)] For integers $n_1,n_2\ge 1$ with $a=n_1+n_2$, an $(n_1,n_2)$ Half-DFS tree implements any $a$-bit Boolean phase table in $T_{\mathrm{HDFS}}=(2^{n_1}-n_1-1)+(2^{n_2}-n_2-1)$ \gate{AND} gates and $Q_{\mathrm{HDFS}}=(2^{n_1}-n_1-1)+(n_2-1)$ temporary ancillas.
\item[(b)] Under rolling or bilinear ANF with constant-term global sign-flip tracking, \gate{MX} uncomputation of any QROM output cancels measurement phase completely using only Clifford gates and zero deferred phase.
\end{enumerate}
\end{proposition}

\begin{proof}
\proofstep{Part (a): Half-DFS complexity.}
Split the address as $z=(z_{\mathrm{lo}},z_{\mathrm{hi}})$ with $a=n_1+n_2$. Materializing all higher-order subset products $u_A$ ($|A|\ge 2$) of $z_{\mathrm{lo}}$ takes $2^{n_1}-n_1-1$ clean ancillas and \gate{AND} gates (single-variable wires $z_{\mathrm{lo},i}$ are already live; pure-prefix terms apply via \gate{Z}$(u_A)$ and any constant phase via a global sign flip). Depth-first traversal of $z_{\mathrm{hi}}$ reuses a stack of $n_2-1$ ancillas and $2^{n_2}-n_2-1$ \gate{AND} gates: while a suffix monomial $v_B$ ($B\neq\emptyset$) is live on the stack, pure-suffix terms apply via \gate{Z}$(v_B)$ and prefix--suffix cross-terms $u_A v_B$ via \gate{CZ}$(u_A,v_B)$ at zero Toffolis; upon backtracking, $v_B$ clears via \gate{MX} with outcome-conditioned \gate{CZ} between its two live parent wires. Summing both halves gives $T_{\mathrm{HDFS}}$ and $Q_{\mathrm{HDFS}}$ as stated (e.g., $(7,8)$ at $a=15$: $367$ Toffolis, $127$ ancillas; $(6,10)$ at $a=16$: $1{,}070$ Toffolis, $66$ ancillas).

\proofstep{Part (b): Immediate ANF Clifford cleanup.}
Measuring a $d_{\mathrm{out}}$-bit QROM output $f(x)=\bigoplus_{A\subseteq[n]} a_A \prod_{i\in A} x_i$ in the $X$ basis with outcome $u_{\mathrm{out}}\in\{0,1\}^{d_{\mathrm{out}}}$ imparts
\begin{equation}
(-1)^{\langle u_{\mathrm{out}},f(x)\rangle} = (-1)^{\langle u_{\mathrm{out}},a_\emptyset\rangle} \prod_{A\neq\emptyset}(-1)^{\langle u_{\mathrm{out}},a_A\rangle \prod_{i\in A} x_i}.
\end{equation}
Under rolling ANF, all subset-product wires $u_A=\prod_{i\in A}x_i$ are live, so outcome-conditioned \gate{Z}$(u_A)$ ($|A|\ge 1$) and a global sign flip conditioned on $\langle u_{\mathrm{out}},a_\emptyset\rangle$ cancel the phase using only Clifford gates. Under a bilinear partition $[n]=I_1\sqcup I_2$ (used in bilinear ANF and reverse Stage~(iii)), every subset splits as $A=A_1\sqcup A_2$ and outcome-conditioned \gate{CZ}$(u_{A_1},u_{A_2})$, \gate{Z}$(u_{A_1})$, \gate{Z}$(u_{A_2})$, and the $A=\emptyset$ sign flip likewise cancel the phase at zero Toffolis.
\end{proof}

\needspace{5\baselineskip}
\subsection{Signed lane addition without sign-extension wires}
\label{app:prop5_trunc}

\begin{proposition}[Signed lane addition without sign-extension wires]
Adding a $b_1$-bit two's-complement signed word $v_{0\dots b_1-1}$ (sign bit $\sigma=v_{b_1-1}$) into a $b_2$-bit lane $R_{0\dots b_2-1}$ ($b_2>b_1$) via \gate{CX}-conjugated ripple incrementation on bits $b_1-1\dots b_2-1$ is exact and reduces Stage-(i) peak workspace by $b_2-b_1$ qubits across all Stage-(i) schedules.
\end{proposition}

\begin{proof}
Modulo $2^{b_2}$, the signed word decomposes as $v_{\mathrm{signed}} \equiv v_{\mathrm{low}} + \sigma(2^{b_2}-2^{b_1-1})$. A clean-carry adder on the low $b_1-1$ bits emits carry $c\in\{0,1\}$ into the upper $(b_2-b_1+1)$-bit slice $R_{\mathrm{high}}$ spanning bits $b_1-1\dots b_2-1$. On this upper slice, adding $c-\sigma$ modulo $2^{b_2-b_1+1}$ satisfies
\[
\bigl(R_{\mathrm{high}} + c - \sigma\bigr) \bmod 2^{b_2-b_1+1} =
\begin{cases}
\bigl(R_{\mathrm{high}} + c\bigr) \bmod 2^{b_2-b_1+1}, & \sigma = 0,\\[2pt]
\overline{\bigl(\overline{R_{\mathrm{high}}} + (c\oplus\sigma)\bigr)\bmod 2^{b_2-b_1+1}}, & \sigma = 1.
\end{cases}
\]
Conjugating $R_{\mathrm{high}}$ and $c$ by \gate{CX}$(\sigma,\cdot)$ around a single-carry ripple incrementer ($b_2-b_1-1$ clean-target \gate{AND} gates plus one terminal \gate{CCX}, totaling $b_2-b_1$ Toffolis) therefore computes $(R+v_{\mathrm{signed}})\bmod 2^{b_2}$ without allocating $b_2-b_1$ sign-extension wires ($5$--$6$ qubits saved for $b_1=21, b_2\in\{26,27\}$).
\end{proof}

\needspace{5\baselineskip}
\subsection{Fixed-point MGF bounds, Hoeffding bounds, and Stinespring trace distance}
\label{app:mgf}

In each Eker{\aa}--H{\aa}stad run, the uniform input $\ket{\psi_{\mathrm{in}}}=2^{-m_e/2}\sum_x\ket{x}$ splits $x$ into $W$ independent uniform windows $x_0,\dots,x_{W-1}$. Let $\mathcal{I}_p = [\mu_p - H_p,\,\mu_p + H_p)$ denote the nonoverflowing interval for offset unsigned $D$-bit lanes.

\leadin{Fixed-point MGF bounds ($D=26\mathrm{u}$).}
For compressed $D=26\mathrm{u}$ lanes, Markov's inequality yields the Chernoff upper-tail bound
\[
\Pr[S_p \ge \mu_p + H_p] \le e^{-\lambda(\mu_p+H_p)} \prod_{j=0}^{W-1} \mathbb{E}\bigl[e^{\lambda d_{p,j}(x_j)}\bigr]
\]
for rational tilt $\lambda=a_\lambda/2^{24}>0$ (and analogously for the lower tail). A float-free C++ worker enumerates all $2^{w_j}$ discrete logarithms per prime and window using 64-bit integer moments rounded upward at fixed-point scale $2^{39}$ ($+1$ unit $2^{-39}$ per rounding step, zero IEEE-754 arithmetic), and an independent verifier lower-bounds Chernoff denominators by summing the 512-term positive Taylor polynomial downward with $256$ fractional bits on a $2^{-192}$ probability grid.

\leadin{Exact rational Hoeffding bounds ($D=27\mathrm{u}$).}
For $D=27\mathrm{u}$ lanes, analytic Hoeffding bounds already suffice. Because $\mu_p=\lfloor\mathbb{E}[S_p]\rceil$ rounds by at most $\tfrac{1}{2}$ and $d_{p,j}\in[0,p-2]$,
\begin{equation}
p_{\mathrm{bad}} \le \sum_{p\in\mathcal{P}} 2\exp\!\left(-\frac{2(H_p - \tfrac{1}{2})^2}{W(p-2)^2}\right),
\label{eq:app_hoeff}
\end{equation}
evaluated conservatively in exact rational Taylor arithmetic over width-256 prime buckets (yielding, for example, $p_{\mathrm{bad}}\le 7.59\times 10^{-15}$ and $\eta \le 1.75\times 10^{-7}$ at $(s=3,\ell=20,w=9,D=27\mathrm{u})$).

\leadin{Stinespring dilation and coherent trace distance.}
Let $\mathcal{G}=\{x\in\{0,1\}^{m_e}:\forall p\in\mathcal{P},\; S_p(x)\in\mathcal{I}_p\}$ with projector $\Pi_{\mathcal{G}}$ on $\mathcal{H}_x$, $\ket{\Psi_{\mathrm{in}}}=\ket{\psi_{\mathrm{in}}}\otimes\ket{\phi_{\mathrm{mask}}}$, and $\Psi_{\mathrm{in}}=\ket{\Psi_{\mathrm{in}}}\bra{\Psi_{\mathrm{in}}}$. Compare the width-$D$ circuit channel $\mathcal{E}_D$ against an uncompressed width-$D'$ reference $\mathcal{E}_{D'}$ with congruent offset $B'_p\equiv B_p\pmod{p-1}$ after tracing out the \gate{MX} environment and reset $\ket{0}$ work registers. Because Propositions~1--5 and Appendix~\ref{app:kickmix_theory} guarantee clean ancilla reset and input-independent phase cancellation on both $\mathcal{G}$ and $\mathcal{G}^c$, Stinespring dilations factor as $\widetilde{V}_d\ket{\Psi}=(U_d\ket{\Psi})\otimes\ket{\phi_d}$ for $d\in\{D,D'\}$. Tracing out the decoupled uniform environment $\ket{\phi_d}$ leaves coherent unitaries agreeing on $\mathcal{G}$, i.e., $(U_D-U_{D'})(\Pi_{\mathcal{G}}\otimes I_Z)=0$, and applying the triangle inequality on $\mathcal{G}^c$ gives
\begin{equation}
\frac{1}{2}\bigl\|\mathcal{E}_D(\Psi_{\mathrm{in}})-\mathcal{E}_{D'}(\Psi_{\mathrm{in}})\bigr\|_1 \le \bigl\|(U_D-U_{D'})(\Pi_{\mathcal{G}^c}\otimes I_Z)\ket{\Psi_{\mathrm{in}}}\bigr\|_2 \le 2\sqrt{p_{\mathrm{bad}}} \le \eta.
\label{eq:app_trace_bound}
\end{equation}
For our primary $s=4,w=8,D=26\mathrm{u}$ family, the integer MGF certificate yields $p_{\mathrm{bad}}\le 2.83\times 10^{-10}$ and $\eta\le 3.37\times 10^{-5}$. Exact per-circuit $\eta$ bounds across all \CmbNumPareto{} Pareto circuits are listed in Table~\ref{tab:cmb_full}.

\clearpage
\section{Complete Compiled Pareto Frontier}
\label{app:tables}

This appendix compiles the full \CmbNumPareto-point Pareto frontier generated by our deterministic bit-level compiler across $\CmbMinQ$--$\CmbMaxQ$ logical qubits under $k\le 24$ (Table~\ref{tab:cmb_full}). All configurations use $\ell=20$-bit primes, offset unsigned lanes $D\mathrm{u}$, 3-bit-sliced Stage~(iii), and fused 3-bit dyadic CRT ($M_{\mathrm{CRT}}=8|\mathcal{P}|$). Low-workspace $w=6$ schedules pair clean-carry adders in Stage~(i) with 2-ancilla clean-\gate{AND} Cuccaro adders in Stages~(iii)--(iv) and $n_1=6$ Half-DFS; wider schedules sweep $b_0\in\{9,10\}$ and copy-duplication factors $(\kappa_{\mathrm{start}},\kappa_{\mathrm{stage}})\in\{(2,1),(4,2),(8,2)\}$.

\leadin{Verification-tier labels in Table~\ref{tab:cmb_full}.}
\emph{All-P $32\times$} denotes dual-backend certification across all primes and $32$ trajectories; \emph{All-P $+$ Dual} denotes exact monolithic counting across all primes paired with dual-backend verification on batch~0; \emph{Dual} denotes dual-backend verification on batch~0 plus 5-batch stratified projection; and \emph{Strat.}\ denotes 5-batch stratified projection.

\begin{table}[!htbp]
\caption{Complete Pareto frontier of \CmbNumPareto{} RSA-2048 circuits compiled at the bit level under $k\le 24$ (\CmbNumParetoEqBudget{} circuits at $f=33$ and $2$ wide-window endpoints at $f=34$). Boldface marks the global minimum qubit--Toffoli product at $2{,}433$ qubits (exact strict product $2.4359\times 10^{12}$ vs.\ $2.4361\times 10^{12}$ at $2{,}381$ qubits; matched $b-2$ product $2.3803\times 10^{12}$ vs.\ $2.3817\times 10^{12}$).}
\label{tab:cmb_full}
\centering
\resizebox{\textwidth}{!}{\footnotesize\renewcommand{\arraystretch}{0.86}\setlength{\tabcolsep}{3.5pt}\begin{tabular}{@{}lcrrrrrrrrl@{}}
\toprule
Stage-(i) schedule, $b_0$ & $(s,k,w,D,f)$ & $|\mathcal{P}|$ & $\Qp$ & $\Ts$ ($10^6$) & $\Tf$ ($10^9$) & $\Tf^{\mathrm{SA}}$ ($10^9$) & $\Qp\Tf$ ($10^{12}$) & Toff./prime & Tail $\eta$ & Check \\
\midrule
Bilinear ANF, low-ws., $b_0{=}8$ & $(14,1,6,26\mathrm{u},33)$ & $20{,}288$ & $1{,}306$ & $417.84$ & $6.3548$ & $6.2943$ & $8.299$ & $20{,}596$ & MGF $7.07\times 10^{-5}$ & All-P $32\times$ \\
Rolling ANF, low-ws., $b_0{=}8$ & $(14,2,6,26\mathrm{u},33)$ & $20{,}288$ & $1{,}359$ & $305.30$ & $4.6432$ & $4.5828$ & $6.310$ & $15{,}049$ & MGF $7.07\times 10^{-5}$ & Dual \\
Rolling ANF, low-ws., $b_0{=}8$ & $(14,3,6,26\mathrm{u},33)$ & $20{,}288$ & $1{,}385$ & $267.44$ & $4.0673$ & $4.0069$ & $5.633$ & $13{,}182$ & MGF $7.07\times 10^{-5}$ & Strat. \\
Rolling ANF, low-ws., $b_0{=}8$ & $(14,4,6,26\mathrm{u},33)$ & $20{,}288$ & $1{,}411$ & $248.85$ & $3.7847$ & $3.7242$ & $5.340$ & $12{,}266$ & MGF $7.07\times 10^{-5}$ & Strat. \\
Rolling ANF, low-ws., $b_0{=}8$ & $(14,6,6,26\mathrm{u},33)$ & $20{,}288$ & $1{,}463$ & $230.07$ & $3.4990$ & $3.4385$ & $5.119$ & $11{,}340$ & MGF $7.07\times 10^{-5}$ & Strat. \\
Zero-tail packed, $b_0{=}9$ & $(8,3,8,27\mathrm{u},33)$ & $16{,}531$ & $1{,}521$ & $340.15$ & $3.0977$ & $3.0684$ & $4.712$ & $20{,}576$ & Hoeff. $2.81\times 10^{-9}$ & Strat. \\
Zero-tail packed, $b_0{=}9$ & $(8,4,8,27\mathrm{u},33)$ & $16{,}531$ & $1{,}548$ & $281.97$ & $2.5679$ & $2.5385$ & $3.975$ & $17{,}057$ & Hoeff. $2.81\times 10^{-9}$ & Strat. \\
Zero-tail packed, $b_0{=}9$ & $(8,5,8,27\mathrm{u},33)$ & $16{,}531$ & $1{,}578$ & $249.36$ & $2.2710$ & $2.2416$ & $3.584$ & $15{,}085$ & Hoeff. $2.81\times 10^{-9}$ & Strat. \\
Zero-tail packed, $b_0{=}9$ & $(8,6,8,27\mathrm{u},33)$ & $16{,}531$ & $1{,}626$ & $227.58$ & $2.0726$ & $2.0432$ & $3.370$ & $13{,}767$ & Hoeff. $2.81\times 10^{-9}$ & Strat. \\
Zero-tail packed, $b_0{=}9$ & $(8,7,8,27\mathrm{u},33)$ & $16{,}531$ & $1{,}674$ & $212.12$ & $1.9318$ & $1.9024$ & $3.234$ & $12{,}832$ & Hoeff. $2.81\times 10^{-9}$ & Strat. \\
Zero-tail packed, $b_0{=}9$ & $(6,6,8,26\mathrm{u},33)$ & $17{,}679$ & $1{,}705$ & $252.29$ & $1.7876$ & $1.7618$ & $3.048$ & $14{,}271$ & MGF $3.34\times 10^{-6}$ & Strat. \\
Zero-tail packed, $b_0{=}9$ & $(6,7,8,26\mathrm{u},33)$ & $17{,}679$ & $1{,}752$ & $234.66$ & $1.6627$ & $1.6369$ & $2.913$ & $13{,}273$ & MGF $3.34\times 10^{-6}$ & Strat. \\
Zero-tail packed, $b_0{=}10$ & $(6,8,8,26\mathrm{u},33)$ & $17{,}679$ & $1{,}861$ & $222.09$ & $1.5736$ & $1.5478$ & $2.929$ & $12{,}563$ & MGF $3.34\times 10^{-6}$ & Strat. \\
Zero-tail packed, $b_0{=}9$ & $(6,10,8,26\mathrm{u},33)$ & $17{,}679$ & $1{,}893$ & $202.82$ & $1.4371$ & $1.4113$ & $2.720$ & $11{,}473$ & MGF $3.34\times 10^{-6}$ & Strat. \\
Rolling ANF, $b_0{=}10$ & $(6,12,8,26\mathrm{u},33)$ & $17{,}679$ & $2{,}003$ & $190.31$ & $1.3484$ & $1.3226$ & $2.701$ & $10{,}764$ & MGF $3.34\times 10^{-6}$ & Strat. \\
Zero-tail packed, $b_0{=}10$ & $(5,11,8,26\mathrm{u},33)$ & $18{,}621$ & $2{,}008$ & $214.23$ & $1.3014$ & $1.2771$ & $2.613$ & $11{,}505$ & MGF $9.04\times 10^{-6}$ & Strat. \\
Rolling ANF, $b_0{=}9$ & $(6,14,8,26\mathrm{u},33)$ & $17{,}679$ & $2{,}055$ & $181.60$ & $1.2867$ & $1.2609$ & $2.644$ & $10{,}272$ & MGF $3.34\times 10^{-6}$ & Strat. \\
Zero-tail packed, $b_0{=}10$ & $(4,10,8,26\mathrm{u},33)$ & $19{,}878$ & $2{,}083$ & $249.02$ & $1.2611$ & $1.2383$ & $2.627$ & $12{,}527$ & MGF $3.37\times 10^{-5}$ & Strat. \\
Rolling ANF, $b_0{=}10$ & $(6,16,8,26\mathrm{u},33)$ & $17{,}679$ & $2{,}107$ & $174.81$ & $1.2386$ & $1.2128$ & $2.610$ & $9{,}888$ & MGF $3.34\times 10^{-6}$ & Strat. \\
Rolling ANF, $b_0{=}9$ & $(5,14,8,26\mathrm{u},33)$ & $18{,}621$ & $2{,}123$ & $198.33$ & $1.2048$ & $1.1805$ & $2.558$ & $10{,}651$ & MGF $9.04\times 10^{-6}$ & Strat. \\
Rolling ANF, $b_0{=}10$ & $(5,16,8,26\mathrm{u},33)$ & $18{,}621$ & $2{,}175$ & $190.89$ & $1.1596$ & $1.1353$ & $2.522$ & $10{,}251$ & MGF $9.04\times 10^{-6}$ & Strat. \\
Rolling ANF, $b_0{=}9$ & $(4,13,8,26\mathrm{u},33)$ & $19{,}878$ & $2{,}199$ & $227.41$ & $1.1517$ & $1.1289$ & $2.533$ & $11{,}441$ & MGF $3.37\times 10^{-5}$ & All-P $32\times$ \\
Rolling ANF, $b_0{=}9$ & $(5,18,8,26\mathrm{u},33)$ & $18{,}621$ & $2{,}227$ & $185.32$ & $1.1257$ & $1.1014$ & $2.507$ & $9{,}952$ & MGF $9.04\times 10^{-6}$ & Strat. \\
Rolling ANF, $b_0{=}10$ & $(4,16,8,26\mathrm{u},33)$ & $19{,}878$ & $2{,}277$ & $213.74$ & $1.0825$ & $1.0596$ & $2.465$ & $10{,}753$ & MGF $3.37\times 10^{-5}$ & Strat. \\
Rolling ANF, $b_0{=}9$ & $(4,20,8,26\mathrm{u},33)$ & $19{,}878$ & $2{,}381$ & $202.03$ & $1.0232$ & $1.0003$ & $2.436$ & $10{,}163$ & MGF $3.37\times 10^{-5}$ & All-P $32\times$ \\
Rolling ANF, $b_0{=}10$ & $(4,22,8,26\mathrm{u},33)$ & $19{,}878$ & $\mathbf{2{,}433}$ & $197.69$ & $\mathbf{1.0012}$ & $\mathbf{0.9783}$ & $\mathbf{2.436}$ & $9{,}945$ & MGF $3.37\times 10^{-5}$ & All-P $32\times$ \\
Rolling ANF, $b_0{=}9$ & $(4,24,8,26\mathrm{u},33)$ & $19{,}878$ & $2{,}485$ & $194.22$ & $0.9836$ & $0.9608$ & $2.444$ & $9{,}771$ & MGF $3.37\times 10^{-5}$ & All-P $32\times$ \\
Zero-tail packed, $b_0{=}9$ & $(3,22,9,27\mathrm{u},33)$ & $19{,}682$ & $2{,}822$ & $241.92$ & $0.9799$ & $0.9620$ & $2.765$ & $12{,}291$ & Hoeff. $1.75\times 10^{-7}$ & All-P + Dual \\
Zero-tail packed, $b_0{=}9$ & $(3,22,9,27\mathrm{u},34)$ & $19{,}682$ & $2{,}823$ & $242.08$ & $0.9782$ & $0.9604$ & $2.762$ & $12{,}299$ & Hoeff. $1.75\times 10^{-7}$ & All-P $32\times$ \\
Zero-tail packed, $b_0{=}10$ & $(3,24,9,27\mathrm{u},34)$ & $19{,}682$ & $2{,}919$ & $234.98$ & $0.9496$ & $0.9317$ & $2.772$ & $11{,}939$ & Hoeff. $1.75\times 10^{-7}$ & All-P + Dual \\
\bottomrule
\end{tabular}
}
\end{table}

\FloatBarrier
\clearpage
\section{Convention Sensitivity, External Comparisons, and Hardware Extensions}
\label{app:external}

This appendix evaluates the robustness of our compiled frontier across alternative cost conventions, concurrent literature benchmarks, and surface-code hardware gate profiles. Appendix~\ref{app:sens} quantifies accounting sensitivity against Gidney's simulated and analytic models (Fig.~\ref{fig:acc}(a)); Appendix~\ref{app:sharma} compares our strict bit-level frontier against the concurrent work of Sharma and Lau~\cite{sharma2026} (Fig.~\ref{fig:acc}(b)); and Appendix~\ref{app:hw_ext} reports exact all-prime hardware gate counts and unconstrained $k>24$ rolling-ANF scaling (Table~\ref{tab:hw_ext}).

\needspace{5\baselineskip}
\subsection{Accounting sensitivity against Gidney (2025)}
\label{app:sens}

Comparing the strict emitted bit-level Toffoli column $\Tf$ against the Gidney-matched $b-2$ column $\Tf^{\mathrm{SA}}$ in Table~\ref{tab:cmb_full} (Eq.~\eqref{eq:app_sa}) confirms that across all \CmbNumPareto{} Pareto circuits, strict $\Tf$ exceeds $\Tf^{\mathrm{SA}}$ by only $0.96\%$--$2.38\%$ ($0.96\%$ at $1{,}306$ qubits, $2.34\%$ at $2{,}433$ qubits, $2.38\%$ at $2{,}485$ qubits, and $1.92\%$ at $2{,}919$ qubits). Thus both the $\sim 70\%$ qubit--Toffoli product reduction and sub-billion factoring hold under either convention (under $\Tf^{\mathrm{SA}}$ alone, the $f=33$ $2{,}822$-qubit control is slightly dominated by the $2{,}485$-qubit circuit).

Figure~\ref{fig:acc}(a) evaluates accounting sensitivity against both our 28 simulated Gidney baselines and Gidney's~\cite{gidney2025} closed-form analytic cost formulas at fixed $(f=33,\mb=26)$ ($\min \Qp\Tf = 7.927\times 10^{12}$ at $1{,}655$ qubits) and with freely retuned $(f,\mb)$ over $f\in[24,59]$ ($\min \Qp\Tf = 7.894\times 10^{12}$ at $1{,}661$ qubits, $f=36$, $\mb=27$), while leaving our compiled frontier unretuned. Because Gidney's closed-form register formula omits $36$ temporary clean ancillas captured by our simulator, analytic comparison is conservative in the baseline's favor. Across all simulated and analytic checks, our minimum qubit--Toffoli product reduction remains $69.85\%$--$70.08\%$ under $\Tf^{\mathrm{SA}}$ and $69.14\%$--$69.39\%$ under strict $\Tf$.

\needspace{5\baselineskip}
\subsection{Comparison with concurrent work (Sharma and Lau, 2026)}
\label{app:sharma}

\begin{figure}[!hbp]
\centering
\includegraphics[width=0.76\textwidth]{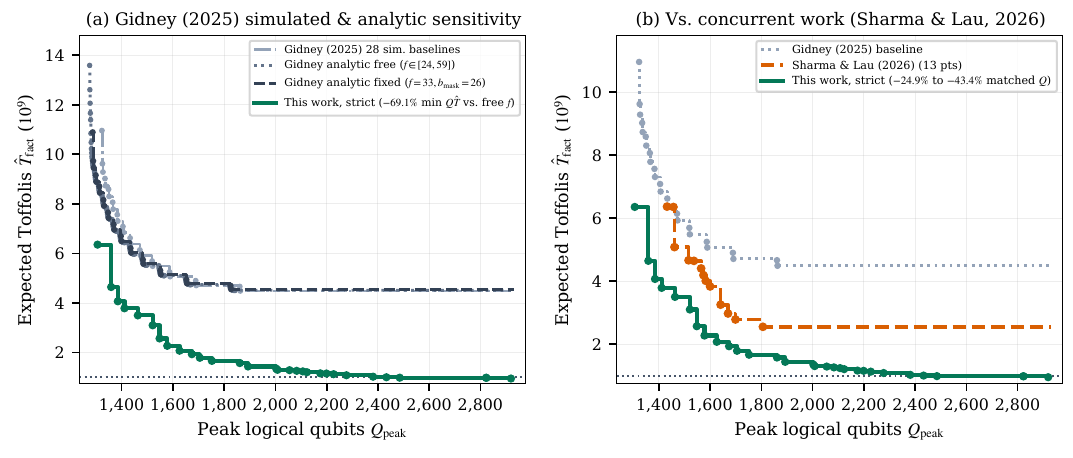}
\caption{External sensitivity and concurrent-work comparisons under strict emitted bit-level Toffoli counting $\Tf$. (a)~Comparison against Gidney's~\cite{gidney2025} 28 simulated baselines and closed-form analytic formulas at fixed $f=33$ and freely retuned $(f,\mb)$ over $f\in[24,59]$. (b)~Comparison against Gidney's baseline and all $13$ published RSA-2048 coordinates of Sharma and Lau~\cite{sharma2026} (flat horizontal extensions mark baseline gate floors).}
\label{fig:acc}
\end{figure}

Figure~\ref{fig:acc}(b) compares our strict bit-level frontier against all $13$ RSA-2048 $(Q,\Tf)$ coordinates ($1{,}432$--$1{,}806$ qubits, fixed $s=8$, omitting the $0.99$ lattice factor) reported in concurrent work by Sharma and Lau~\cite{sharma2026}. Reference~\cite{sharma2026} reports macro-level Toffoli estimates without published bit-level streams, making direct comparison against our strict bit-level counts conservative in favor of Ref.~\cite{sharma2026}.

Across every reported budget ($1{,}432$--$1{,}806$ qubits), our circuits save $24.93\%$--$43.36\%$ expected Toffolis at equal or lower qubit count ($30.26\%$--$43.36\%$ across matched $s=8$ points from $1{,}521$ to $1{,}674$ qubits). Relative to their lowest reported $2.5492\times 10^9$ floor at $1{,}806$ qubits, our $2{,}919$-qubit $f=34$ endpoint is $62.75\%$ lower under strict bit-level counting ($63.45\%$ under $\Tf^{\mathrm{SA}}$) using $1{,}113$ additional qubits.

\needspace{12\baselineskip}
\subsection{Exact all-prime hardware gate profiles and unconstrained \texorpdfstring{$k>24$}{k > 24} scaling}
\label{app:hw_ext}

\leadin{Exact all-prime opcode profiles.}
Table~\ref{tab:hw_ext}(a) reports exact monolithic opcode breakdowns across all constituent primes for the six certified circuits, separating non-Clifford \gate{AND}/\gate{CCX}/\gate{CSWAP} counts from explicit two-qubit Cliffords (\gate{CX}, \gate{CZ}) and total $X$-basis \gate{HMR} operations (explicit \gate{MX} plus \gate{AND} uncomputations), directly supporting the surface-code tradeoffs in Sec.~\ref{sec:discussion}. In particular, widening the Stage-(iii) startup lookup from $b_0=9$ to $b_0=10$ at $2{,}433$ qubits saves only $0.04\%$ net Toffolis yet doubles explicit conditional \gate{CZ} volume from $0.65\times 10^9$ to $1.30\times 10^9$.

\leadin{Unconstrained $k>24$ rolling-ANF scaling.}
Table~\ref{tab:hw_ext}(b) documents how rolling ANF scales when the main-frontier batch cap $k\le 24$ is relaxed to $k\in\{26,30,36,40,42\}$, while staying strictly within Gidney's $f=33$ error budget and keeping zero segmentation overhead. At $k=42$ and $2{,}953$ qubits, expected factoring cost reaches $\Tf^{\mathrm{SA}}=0.8757\times 10^9$ (strict bit-level $\Tf=0.8986\times 10^9$).

\begin{table}[!htbp]
\caption{Exact monolithic hardware gate profiles across all primes for the six certified circuits, together with the unconstrained $k>24$ rolling-ANF extension. Part~(a) counts explicit \gate{CX}$+$\gate{CZ} (unfolding \gate{AND} releases adds $1$ conditional \gate{CZ} and \gate{CSWAP} (\gate{CSW}) adds $2$ \gate{CX}, at most $+1.1$ two-qubit Cliffords per Toffoli) alongside total \gate{HMR} operations (explicit \gate{MX} plus \gate{AND} releases).}
\label{tab:hw_ext}
\centering
\resizebox{0.94\textwidth}{!}{\footnotesize\renewcommand{\arraystretch}{0.88}\setlength{\tabcolsep}{4.5pt}\begin{tabular}{@{}lcrrrrrrr@{}}
\toprule
\multicolumn{9}{@{}l}{\textbf{(a) Exact all-prime monolithic non-Clifford, 2Q Clifford, and measurement counts across all six certified circuits}} \\
\midrule
Schedule, $b_0$ & $(s,k,w,f)$ & $\Qp$ & Toff./run ($10^6$) & \gate{AND}/\gate{CCX}/\gate{CSW} (\%) & \gate{CX} ($10^9$) & \gate{CZ} ($10^9$) & 2Q Cliff./Toff. & \gate{HMR} ($10^6$) \\
\midrule
Bilinear ANF, low-ws., $b_0{=}8$ & $(14,1,6,33)$ & $1{,}306$ & $417.84$ & $99.0 / 1.0 / 0.1$ & $3.75$ & $1.52$ & $12.6\times$ & $512.6$ \\
Rolling ANF, $b_0{=}9$ & $(4,13,8,33)$ & $2{,}199$ & $227.41$ & $95.5 / 1.7 / 2.8$ & $11.27$ & $0.65$ & $52.4\times$ & $322.3$ \\
Rolling ANF, $b_0{=}9$ & $(4,20,8,33)$ & $2{,}381$ & $202.03$ & $95.0 / 1.9 / 3.1$ & $11.27$ & $0.65$ & $59.0\times$ & $297.0$ \\
Rolling ANF, $b_0{=}10$, min $\Qp\Tf$ & $(4,22,8,33)$ & $2{,}433$ & $197.69$ & $94.0 / 1.9 / 4.0$ & $11.30$ & $1.30$ & $63.7\times$ & $292.6$ \\
Rolling ANF, $b_0{=}9$, strict $<10^9$ & $(4,24,8,33)$ & $2{,}485$ & $194.22$ & $94.8 / 2.0 / 3.3$ & $11.27$ & $0.65$ & $61.4\times$ & $289.2$ \\
Zero-tail packed, $b_0{=}9$ & $(3,22,9,34)$ & $2{,}823$ & $242.08$ & $95.9 / 1.5 / 2.6$ & $21.04$ & $0.65$ & $89.6\times$ & $335.5$ \\
\midrule
\multicolumn{9}{@{}l}{\textbf{(b) Unconstrained $k>24$ rolling-ANF extension at $(s{=}4, \ell{=}20, w{=}8, D{=}26\mathrm{u}, f{=}33)$, zero seg.\ overhead}} \\
\midrule
Schedule & $(s,k,w,f)$ & $\Qp$ & $\Ts$ ($10^6$) & $\Tf^{\mathrm{SA}}$ ($10^9$) & $\Tf$ ($10^9$) & $\Qp\Tf^{\mathrm{SA}}$ ($10^{12}$) & $\Qp\Tf$ ($10^{12}$) & Toff./prime \\
\midrule
Rolling ANF, $k{=}26$ & $(4,26,8,33)$ & $2{,}537$ & $190.97$ & $0.9443$ & $0.9672$ & $2.396$ & $2.454$ & $9{,}607$ \\
Rolling ANF, $k{=}30$ & $(4,30,8,33)$ & $2{,}641$ & $186.22$ & $0.9202$ & $0.9431$ & $2.430$ & $2.491$ & $9{,}368$ \\
Rolling ANF, $k{=}36$ & $(4,36,8,33)$ & $2{,}797$ & $181.01$ & $0.8939$ & $0.9167$ & $2.500$ & $2.564$ & $9{,}106$ \\
Rolling ANF, $k{=}40$ & $(4,40,8,33)$ & $2{,}901$ & $178.43$ & $0.8808$ & $0.9036$ & $2.555$ & $2.621$ & $8{,}976$ \\
Rolling ANF, $k{=}42$ & $(4,42,8,33)$ & $2{,}953$ & $177.43$ & $0.8757$ & $0.8986$ & $2.586$ & $2.654$ & $8{,}926$ \\
\bottomrule
\end{tabular}}
\end{table}

\end{appendix}

\end{document}